\documentclass[10pt, conference, letterpaper]{IEEEtran}

\usepackage{balance}  
\usepackage{graphicx} 
\usepackage{times}    
\usepackage{url}      
\usepackage{cite}
\usepackage{balance}
\usepackage{color}
\usepackage{wrapfig}
\usepackage{caption}
\usepackage{subcaption}
\usepackage{array}
\usepackage{marginnote}
\usepackage{epstopdf}
\usepackage{xcolor}
\usepackage{amsfonts}
\usepackage{flushend}
\usepackage{color}

\def \sys {\textit{WiGest}}

\begin{document}

\title{WiGest: A Ubiquitous WiFi-based Gesture Recognition System} 

\author{\IEEEauthorblockN{Heba Abdelnasser}
\IEEEauthorblockA{Computer and Sys. Eng. Department\\
Alexandria University \\
heba.abdelnasser@alexu.edu.eg}
\and
\IEEEauthorblockN{Moustafa Youssef}
\IEEEauthorblockA{Wireless Research Center\\
Egypt-Japan Univ. of Sc. and Tech.\\
moustafa.youssef@ejust.edu.eg}
\and
\IEEEauthorblockN{Khaled A. Harras}
\IEEEauthorblockA{Computer Science Department\\
Carnegie Mellon University\\
kharras@cs.cmu.edu}
}

\maketitle

\begin{abstract}
We present WiGest: a system that leverages changes in WiFi signal strength to sense in-air hand gestures around the user's mobile device. Compared to related work, WiGest is unique in using standard WiFi equipment, with no modifications, and no training for gesture recognition. The system identifies different signal change primitives, from which we construct mutually independent gesture families. These families can be mapped to distinguishable application actions. We address various challenges including cleaning the noisy signals, gesture type and attributes detection, reducing false positives due to interfering humans, and adapting to changing signal polarity. We implement a proof-of-concept prototype using off-the-shelf laptops and extensively evaluate the system in both an office environment and a typical apartment with standard WiFi access points. Our results show that \sys{} detects the basic primitives with an accuracy of 87.5\% using a single AP only, including through-the-wall non-line-of-sight scenarios. This accuracy increases to 96\% using three overheard APs. In addition, when evaluating the system using a multi-media player application, we achieve a classification accuracy of 96\%. This accuracy is robust to the presence of other interfering humans, highlighting \sys{'s} ability to enable future ubiquitous hands-free gesture-based interaction with mobile devices.

\end{abstract}
\IEEEpeerreviewmaketitle

\section{Introduction}
\label{introduction}
The exponential growth in mobile technologies has reignited the investigation of novel human-computer interfaces (HCI) through which users can control various applications. Motivated by freeing the user from specialized devices and leveraging natural and contextually relevant human movements, gesture recognition systems are becoming increasingly popular as a fundamental approach for providing HCI alternatives. Indeed, there is a rising trend in the adoption of gesture recognition systems into various consumer electronics and mobile devices,  including smart phones \cite{gupta2012soundwave}, laptops \cite{QosmioG55}, navigation devices \cite{RakuNaviGPS}, and gaming consoles \cite{shotton2013real}.
 These systems, along with research enhancing them by exploiting the wide range of sensors available on such devices, generally adopt various techniques for recognizing gestures including computer vision \cite{shotton2013real}, inertial sensors \cite{cohn2012humantenna, harrison2010skinput, kim2012digits}, ultra-sonic \cite{gupta2012soundwave}, and infrared (e.g. on the Samsung S4 phone). While promising, these techniques experience various limitations such as being tailored for specific applications, sensitivity to lighting, high installation and instrumentation overhead, requiring holding the mobile device, and/or requiring additional sensors to be worn or installed.

With the ubiquity of WiFi-enabled devices and infrastructure, WiFi-based gesture recognition systems, e.g. \cite{scholz2011challenges,adib2013see,pu2013whole}, have recently been proposed to help overcome the above limitations in addition to  
 enabling users to provide \textbf{\emph{in-air hands-free}} input to various applications running on mobile devices. This is particulary useful in cases when a user's hands are wet, dirty, busy, or she is wearing gloves; rendering touch input difficult. These WiFi-based systems are based on analyzing the changes in the characteristics of the wireless signals, such as the received signal strength indicator (RSSI) or detailed channel state information (CSI), caused by human motion. However, due to the noisy wireless channel and complex wireless propagation, these systems either require \textbf{calibration} of the area of interest or \textbf{major changes} to the standard WiFi hardware to extract the desired signal features, therefore limiting the adoption of these solutions using off-the-shelf components. Moreover, all these systems 
do not provide \textbf{fine-grained} control of a specific user mobile device. 

\begin{figure}[!t]
\centering
	\begin{subfigure}[t]{0.23\textwidth}
	\centering
	\includegraphics[width=\textwidth]{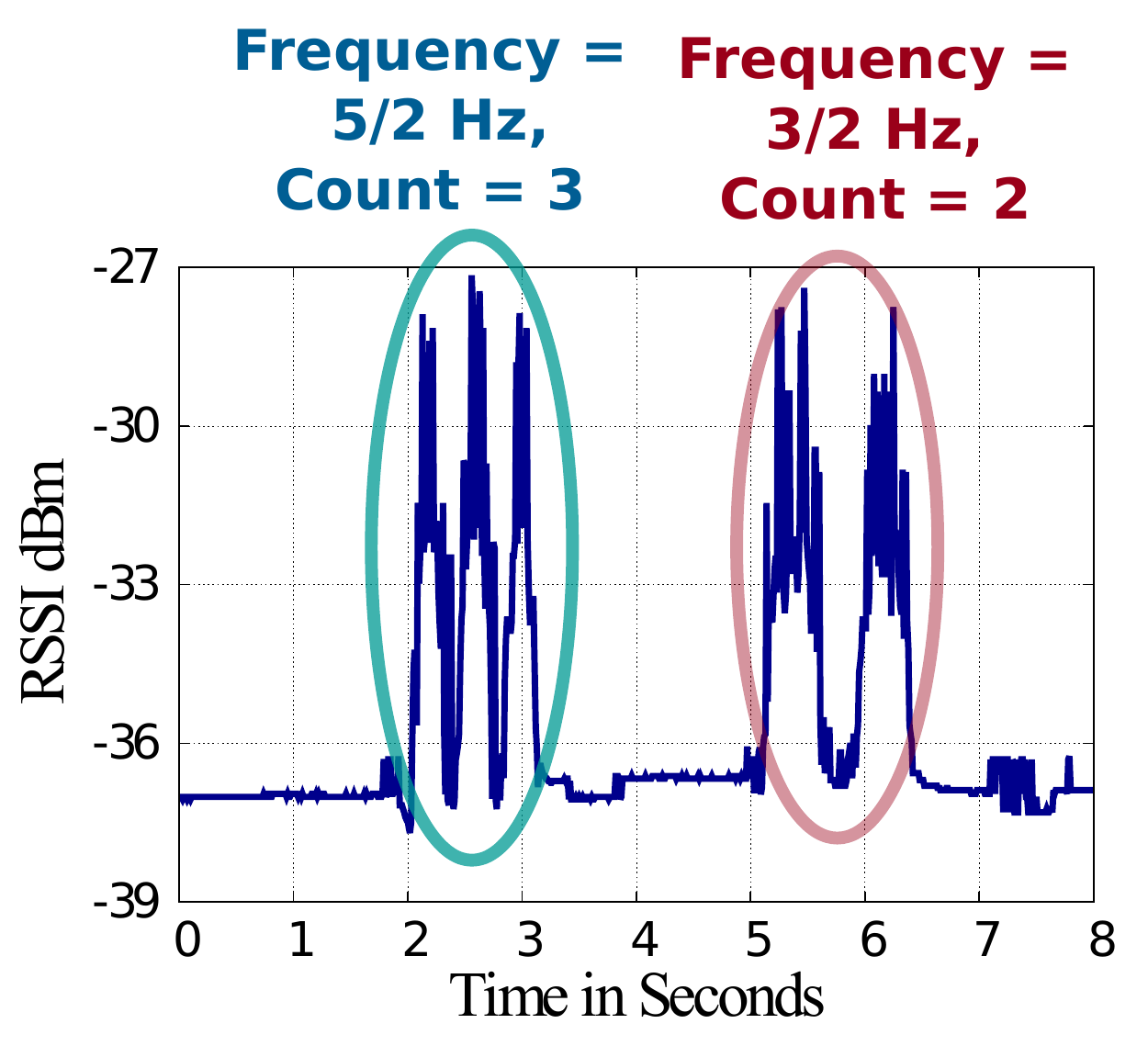}
	\caption{Raw signal}
	\label{fig:intro_raw}
	\end{subfigure}
	\begin{subfigure}[t]{0.23\textwidth}
	\centering
	\includegraphics[width=\textwidth]{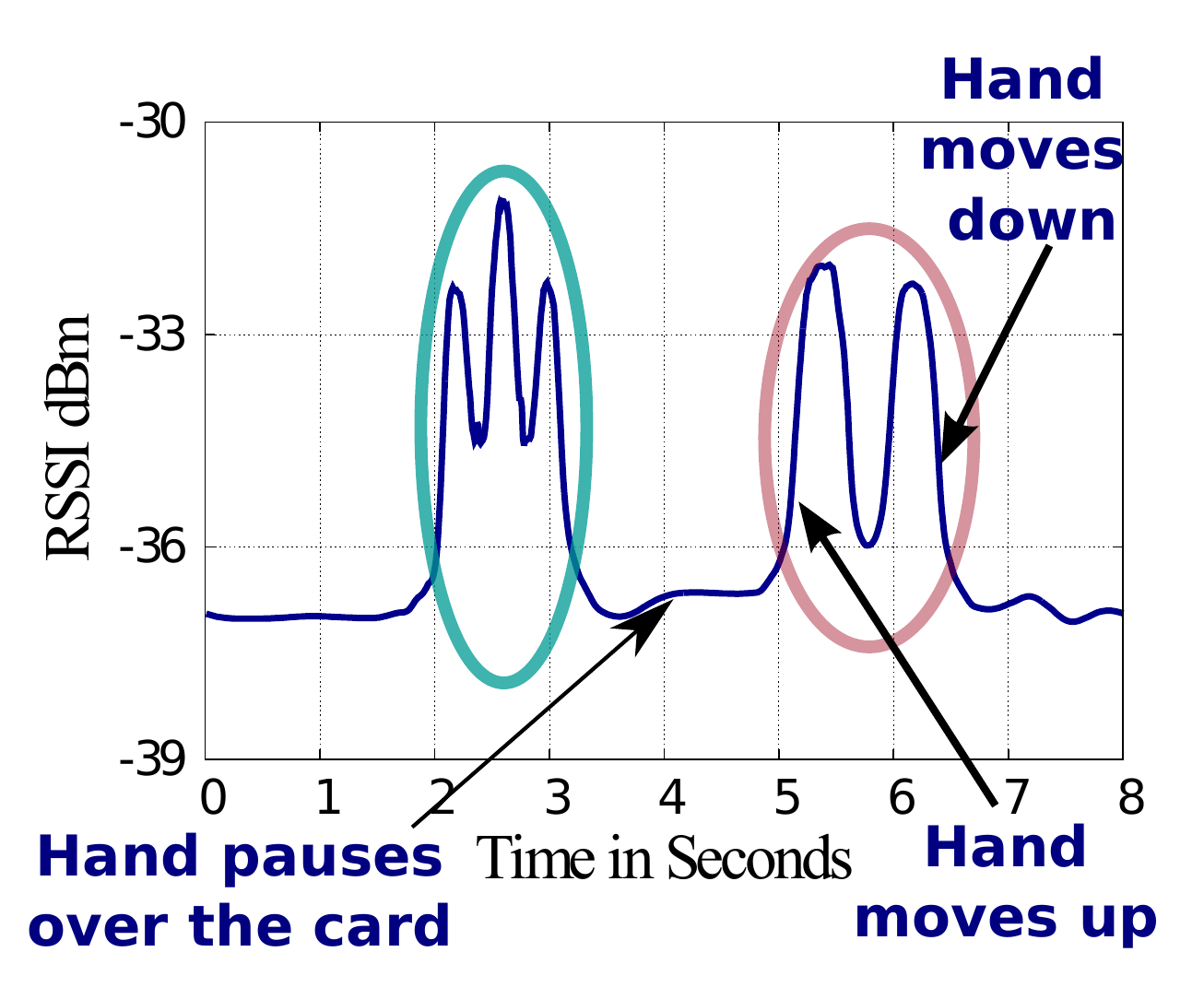}
	\caption{Denoised signal}
	\label{fig:intro_denoised}
	\end{subfigure}
\caption{The impact of two hand motion frequencies on the RSSI. The received signal is composed of three primitives: rising edge, falling edge, and pause. The five motion repetitions can also be counted.}
\vspace{-0.2in}
\label{fig:noisy_gesture}
\end{figure}

This paper presents \sys{}\footnote{\textbf{A video showing WiGest in action can be found at:} \texttt{\url{http://wrc-ejust.org/projects/wigest/}}. \textbf{A demo of WiGest has been presented at the IEEE Infocom 2015 conference \cite{abdelnasser2015wigest}.}}, a ubiquitous WiFi-based hand gesture recognition system for controlling applications running on off-the-shelf WiFi-equipped devices. \sys{} does not require additional sensors, is resilient to changes within the environment, does not require training, and can operate in non-line-of-sight scenarios.
The basic idea is to leverage the effect of the in-air hand motion on the wireless signal strength received by the device from an access point to recognize the performed gesture. Figure \ref{fig:noisy_gesture} demonstrates the impact of some hand motion gestures within proximity of the receiver on the RSSI values, creating three unique signal states (we call primitives): a rising edge, a falling edge, and a pause. \sys{} parses combinations of these primitives along with other parameters, such as the speed and magnitude of each primitive, to detect various gestures. These gestures, in turn, can be mapped to distinguishable application actions.
Since \sys{} works with a specific user device (e.g. a laptop or mobile phone), it has the advantage of removing the ambiguity of the user location relative to the receiver. This helps eliminate the requirement of calibration and special hardware.

There are several challenges, however, that need to be addressed to realize \sys{} including handling the  noisy RSSI values due to multipath interference, medium contention, and other electromagnetic noise in the wireless medium; handling the variations of gestures and their attributes for different humans and even for the same human at different times; handling interference (i.e. avoiding false positives) due to the motion of other people within proximity of the user's device; and finally be energy-efficient to suit mobile devices.

To address these challenges, \sys{} leverages different signal processing techniques that can preserve signal details while filtering out the noise and variations in the signal. In addition, we leverage multiple overheard APs as well as introduce a unique signal pattern (acting as a preamble) to identify the beginning of the gesture engagement phase; this helps counter the effect of interfering humans, increase robustness, and enhance accuracy. Finally, \sys{} energy-efficiency stems from the fact that detecting the preamble can be efficiently done based on a simple thresholding approach, rendering the system idle most of the time. In addition, we use an efficient implementation of the wavelet transform that has a linear time complexity in various signal processing modules.

We implement \sys{} on off-the-shelf laptops and evaluate its performance with three different users in a  1300$\textrm{ft}^{2}$ three bedroom apartment and a 5600$\textrm{ft}^{2}$ floor within our engineering building. Various realistic scenarios are tested covering  more than 1000 primitive actions and gestures each in the presence of interfering users in the same room as well as other people moving in the same floor during their daily life.

Our results reveal that \sys{} can detect the basic primitives with an accuracy of 87.5\% using a single AP only for distances up to 26 \textit{ft} including through-the-wall non-line-of-sight scenarios. This accuracy increases to 96\% using three overheard APs, which is the typical case for many WiFi deployment scenarios. In addition, when evaluating the system using a multi-media player application case study, we achieve a classification accuracy of 96\%. This accuracy is robust to the presence of other interfering humans.

The rest of the paper is organized as follows. Section \ref{related_work} provides an overview on related work. Sections \ref{system_layers} and \ref{system_logic_flow} present an overview and the details of the \sys{} system respectively. We evaluate the system in  Section~\ref{evaluation}. Finally, we conclude the paper and provide directions for future work in Section~\ref{conclusion}.

\begin{figure*}[!t]
\centering
\scalebox{0.9}{
        \begin{subfigure}[b]{0.5\textwidth}
                 \centering
                \includegraphics[width=\textwidth]{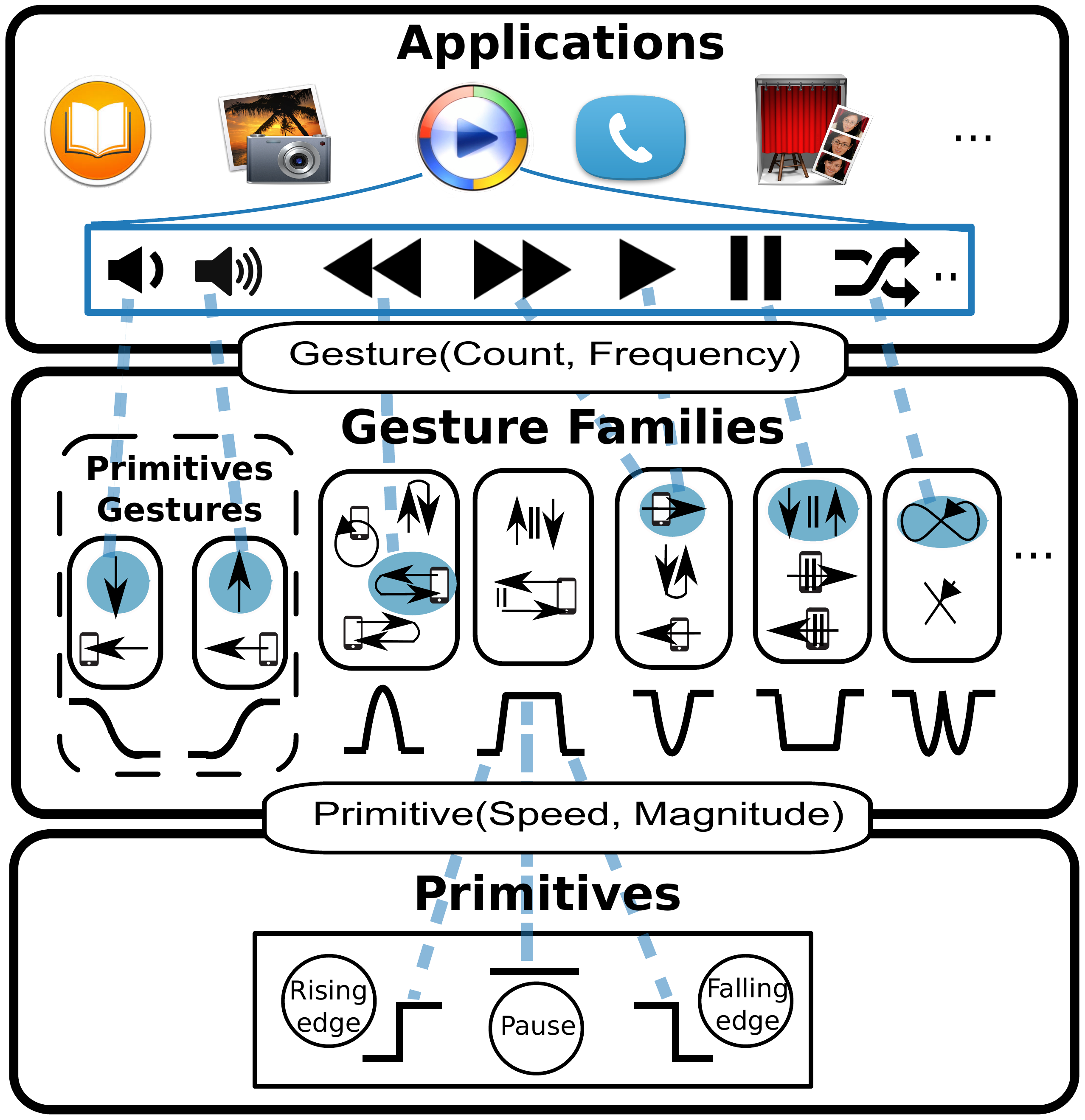}
                \caption{Layers overview.}
                \label{fig:architecture:layers_overview}
        \end{subfigure} ~~~~
        \begin{subfigure}[b]{0.45\textwidth}
        		\centering
		\includegraphics[width=\textwidth]{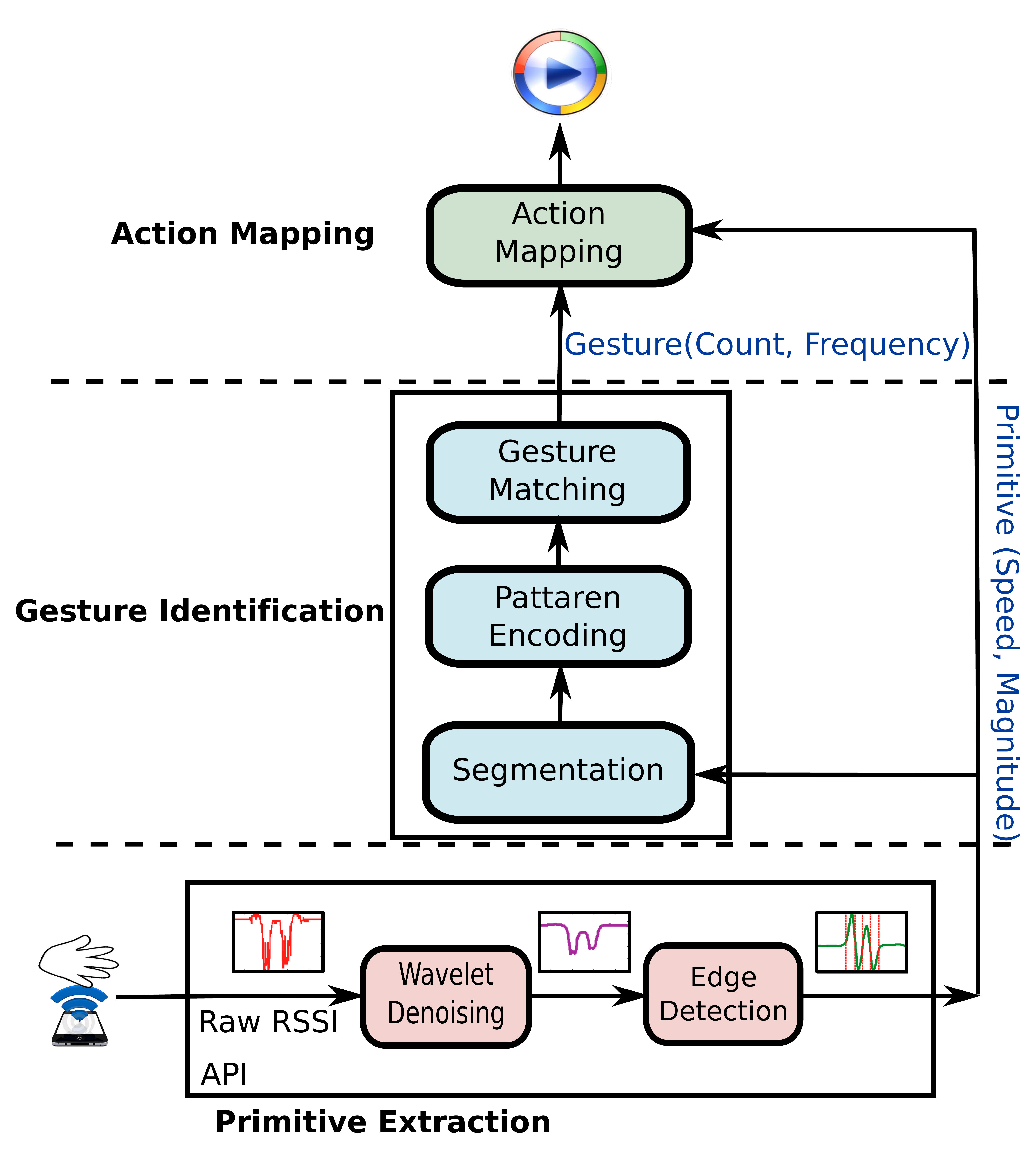}
		\caption{System logic flow.}
		\label{fig:architecture:logic_flow}
        \end{subfigure}
        }
 \caption{\sys{} (a) conceptual view and (b) processing components. \sys{} has three layers: primitives, gestures, and applications. The figure shows a typical example for mapping the application actions to different example gestures and primitives. We note that multiple application actions (e.g. ``play'' and ``fast'' forward) can be mapped to the same gesture with different parameters (e.g., one count for play and more than one count for fast forward).}
 \label{fig:architecture}
 \vspace{-0.1in}
\end{figure*}

\section{Related Work}
\label{related_work}

Gesture recognition systems generally adopt various techniques such as computer vision \cite{shotton2013real}, inertial sensors~\cite{cohn2012humantenna}, ultra-sonic \cite{gupta2012soundwave}, and infrared electromagnetic radiation (e.g. on Samsung S4). While promising, these techniques suffer from limitations such as sensitivity to lighting, high installation and instrumentation overhead, demanding dedicated sensors to be worn or installed, or requiring line-of-sight communication between the user and the sensor. These limitations promoted exploiting WiFi, already installed on most user devices and abundant within infrastructure, for \textit{activity} and \textit{gesture} recognition as detailed in this section.

\subsection{WiFi-based Activity Recognition Systems}
Device-free activity recognition relying on RSSI fluctuations, or the more detailed channel state information (CSI) in standard WiFi networks, have emerged as a ubiquitous solution for presence detection \cite{kosba2012rasid, youssef2007challenges, saeed2014ichnaea, abdel2013monophy,Kafrawy:PIMRC11, aly2013new, seifeldin2011kalman, seifeldin2010deterministic, eleryan2011aroma, kosba2012robust, sabek2012multi,smartdevices}, tracking \cite{xiao2013pilot, kosba2009analysis, el2010propagation, seifeldin2013nuzzer,Sabek:TRSPOT12,sabek2014ace}, and recognition of human activities\cite{ding2011rftraffic,al2012rf,sigg2013rfa, Kassem:VTC12}. For activity recognition systems, RFTraffic \cite{ding2011rftraffic} introduces a system that classifies the traffic density scenarios based on the emitted RF-noise from the vehicles, while \cite{al2012rf,Kassem:VTC12} can further differentiate between vehicles and humans and detect the vehicle speed. In \cite{shi2013joint}, authors provide localization and detect basic human activities such as standing and walking, using ambient FM broadcast signals, or by sensing WiFi RSSI changes within proximity of a mobile device. In addition, activities conducted by multiple individuals can be detected simultaneously with good accuracy utilizing multiple receivers and simple features and classifier systems \cite{sigg2013rfa}, which can be further used in novel application domains, such as automatic indoor semantic identification \cite{alzantot2012crowdinside,checkinside,wang2012no}.

The solutions described above typically require some prior form of training, have been tested in controlled environments, or work in scenarios where human presence is rare (e.g. intrusion detection), and most importantly, do not detect fine-grained motions such as hand gestures near mobile devices. \sys{} builds on this foundational work to achieve fine-grained hand gesture recognition based on RSSI changes \textbf{without requiring} any training and in typical daily scenarios.

\subsection{RF-based Gesture Recognition Systems}

Work in this area represents the most recent efforts to detect fine-grained mobility and human gestures by leveraging RF signals. WiVi \cite{adib2013see} uses an inverse synthetic aperture radar (ISAR) technique by treating the motion of a human body as an antenna array to track the resulting RF beam, thus enabling radar-like vision and simple through-wall gesture-based communication. 
Finally, WiSee \cite{pu2013whole} is a fine-grained gesture recognition system that builds on DopLink \cite{aumi2013doplink} and \cite{kim2009human} by exploiting the doppler shift in narrow bands extracted form wide-band OFDM transmissions to recognize nine different human gesturers.

While these solutions provide high accuracy, they all require \textbf{special hardware} in order to employ their solutions. \sys{}, on the other hand, works with off-the-shelf WiFi components, making it significantly more deployable. Moreover, working with off-the-shelf components requires dealing with more challenges in terms of noise and interference sources.

\section{WiGest Conceptual Overview}
\label{system_layers}

In this section, we provide a conceptual overview of \sys{} covering the Primitives, Gesture Families, and Application Actions layers (Figure~\ref{fig:architecture:layers_overview}).

\subsection{Primitives Layer}
\begin{figure}[!t]
\centering
        \begin{subfigure}[b]{0.23\textwidth}
                 \centering
                \includegraphics[width=\textwidth]{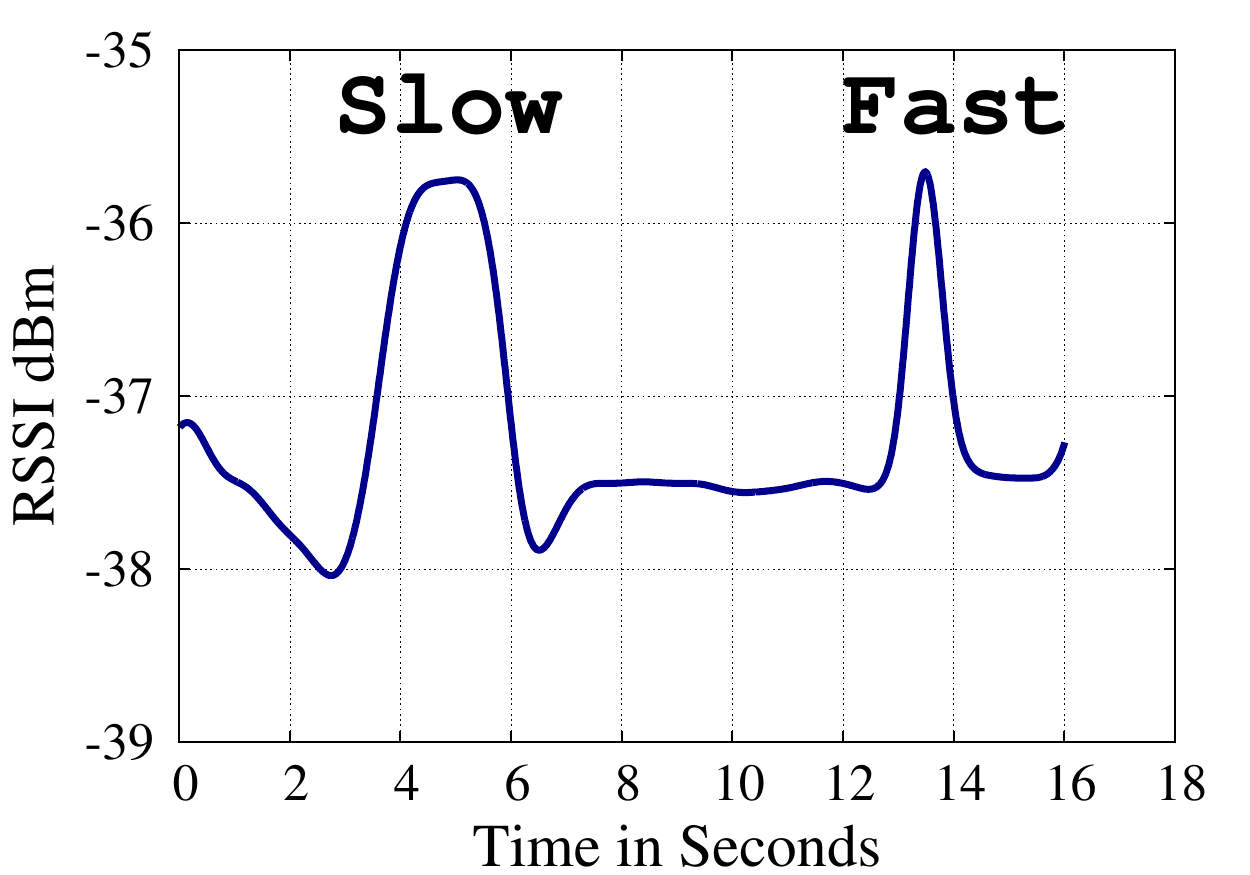}
                \caption{Hand motion speed effect: The slower the speed, the slower the RSSI change.}
                \label{fig:raw_speed}
        \end{subfigure}
        \begin{subfigure}[b]{0.23\textwidth}
        		\centering
		\includegraphics[width=\textwidth]{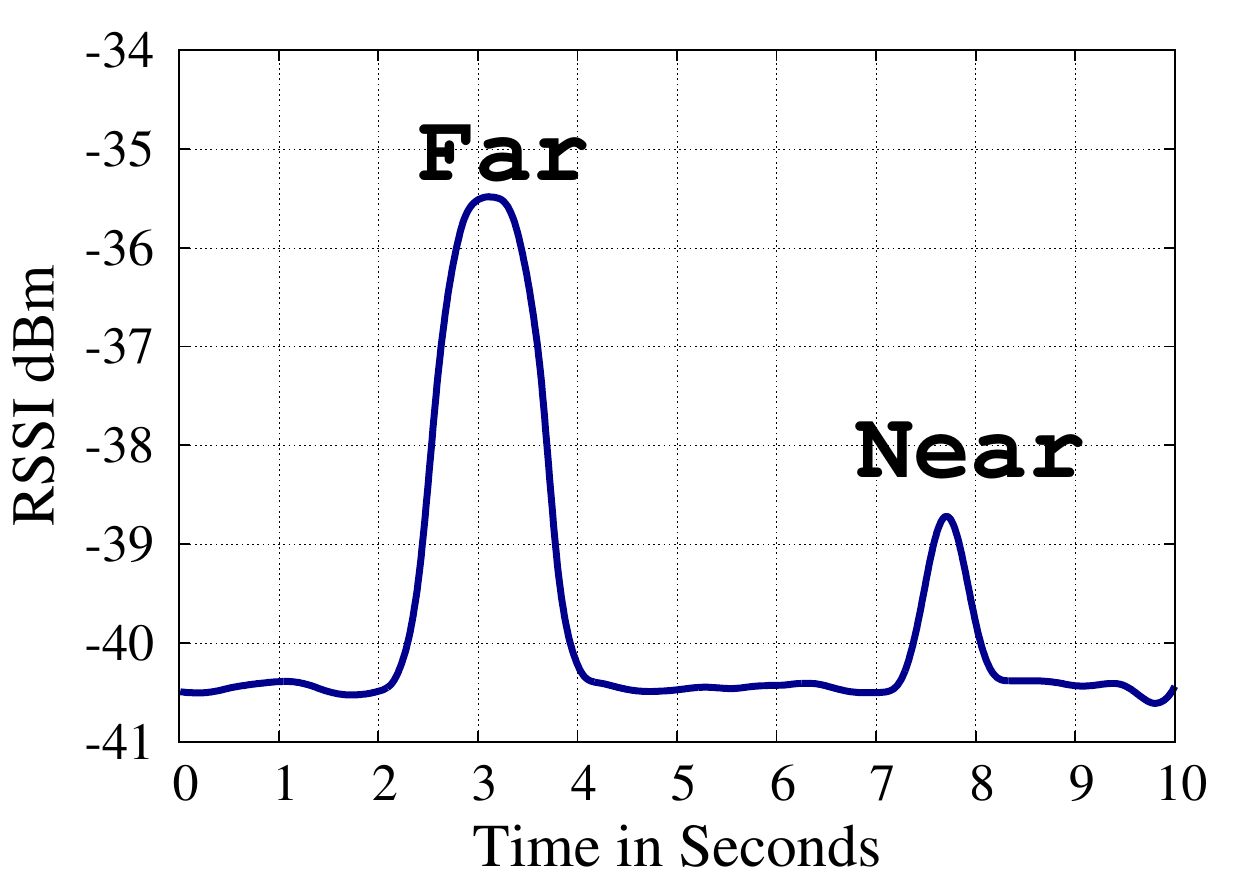}
		\caption{Hand motion distance effect. The higher the distance, the higher the change in RSSI.}
		\label{fig:raw_dist}
        \end{subfigure}
 \caption{Parameters associated with raw signal primitives.}
 \vspace{-0.2in}
 \label{fig:prim_param}
\end{figure}

This layer detects the basic changes in the raw RSSI. These changes include rising edges caused by moving the hand away from the device, falling edges caused by moving the hand towards the device, and pauses caused by holding the hand still over the device. Higher layer gestures can be composed by combining these three primitives. There are other parameters that can be associated with these primitive actions. In particular, for the rising and falling edges, \sys{} can extract the speed of motion and magnitude of signal change (Figure~\ref{fig:prim_param}). To reduce the noise effect, \sys{} discretizes the values of the parameters. Therefore, there are three defined speeds (high, medium, and low) and two defined magnitudes (far and near).

\begin{figure}[!t]
\centering
        \begin{subfigure}[b]{0.11\textwidth}
                \centering
                \includegraphics[width=\textwidth]{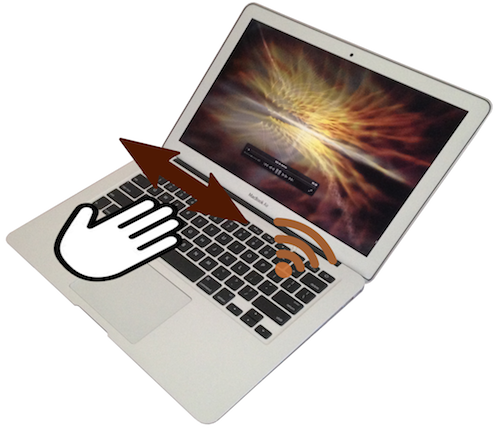}
                \caption{Right-left }
                \label{fig:rl}
        \end{subfigure}
   	    \begin{subfigure}[b]{0.11\textwidth}
                \centering
                \includegraphics[width=\textwidth]{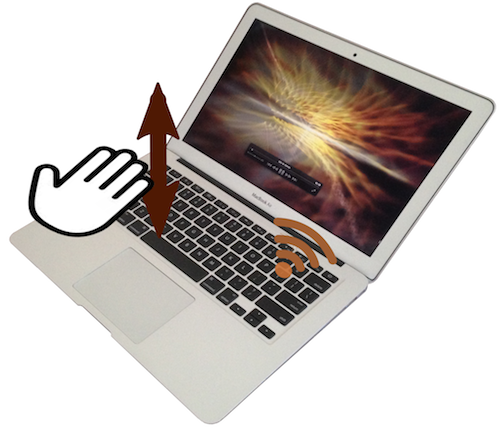}
                \caption{Up-down}
                \label{fig:ud}
        \end{subfigure}
        \begin{subfigure}[b]{0.11\textwidth}
                \centering
                \includegraphics[width=\textwidth]{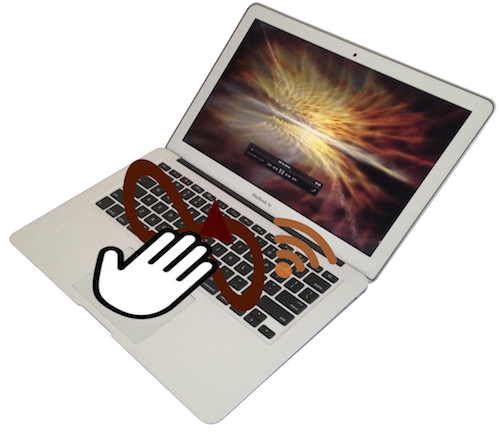}
		\caption{Infinity}
                \label{fig:inf}
        \end{subfigure}
         \begin{subfigure}[b]{0.11\textwidth}
                \centering
                \includegraphics[width=\textwidth]{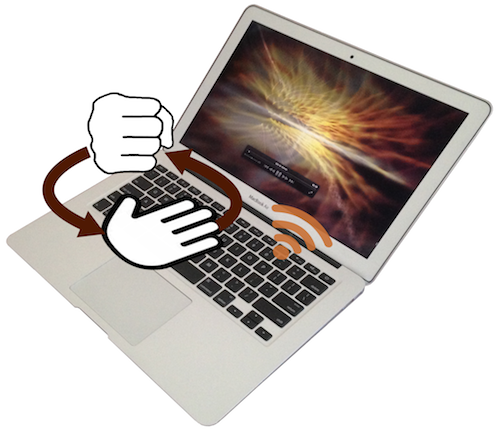}
		\caption{Open-close}
                \label{fig:och}
        \end{subfigure}
	\caption{Some gesture sketches detected by \sys{}.}
    \vspace{-0.2in}
    \label{fig:gestures}
\end{figure}

\subsection{Gesture Families Layer}
Different basic primitives from the lower layer can be combined to produce higher level gestures (Figure~\ref{fig:gestures}). For example, an up-down hand gesture can be mapped to the primitives rising then falling edges. Since there may be ambiguity between different hand gestures, we define the concept of a \emph{gesture family}, which represents a set of gestures that have the same sequence of primitives. For example, as shown in Figure~\ref{fig:architecture:layers_overview},  all up-down, right-left and left-right hand gestures have the same effect on the signal strength and hence the same primitives sequence of a falling then rising edge. Gesture families give flexibility to the developers to choose the best gesture that fits their applications from a certain gesture family. Similar to the basic primitives, gestures can have associated attributes. Specifically, the count of the gesture repetition (e.g. how many consecutive up-down gestures) and the frequency (how fast the repetition is performed) are two attributes that can be associated with the gestures (Figure~\ref{fig:intro_denoised}).

\subsection{Application Actions Layer}
This layer is where each application maps its actions to different gestures. Typically, one action is mapped to one gesture family and the developer can pick one or more gestures from the same family to represent an action. As an example, for a media player application (Figure~\ref{fig:architecture}) a ``play'' action can be performed with a right-movement hand gesture while a ``volume up'' action can be mapped to moving the hand up. The hand movement speed can be mapped to the rate of volume change.

In the next section, we give the details of extracting these different semantics and the associated challenges.

\section{The WiGest System}
\label{system_logic_flow}

In this section, we discuss the processing flow of our system depicted in Figure \ref{fig:architecture:logic_flow} and address the associated challenges. This flow includes the three main stages that map to the different semantic layers: Primitives Extraction, Gesture Identification, and Action Mapping.

\subsection{Primitives Extraction}
The goal of this stage is to extract the basic primitives (rising edges, falling edges, and pauses) from the raw signal. Due to the noisy wireless signal, this stage starts by a noise reduction step using the Discrete Wavelet Transform (DWT), followed by an edge extraction and primitives detection step. We start with a brief background on the DWT. Then we give the details of the different submodules.

\subsubsection{Discrete Wavelet Transform}

Wavelet Transform provides a time-frequency representation of a signal. It has two main advantages: (a) an optimal resolution both in the time and the frequency domains; and (b) it achieves fine-grained multi-scale analysis \cite{Wavlet_book}. In Discrete Wavelet Transform (DWT), the generic step splits the signal into two parts: an approximation coefficient vector and a detail coefficient vector. This splitting is applied recursively a number of steps (i.e. levels), $J$, to the approximation coefficients vector only to obtain finer details from the signal. At the end, DWT produces a coarse approximation projection (scaling) coefficients $\alpha^{(J)}$, together with a sequence of finer detail projection (wavelet) coefficients $\beta^{(1)},  \beta^{(2)}, ...,\beta^{(J)}$. The DWT coefficients in each level can be computed using the following equations:

\begin{equation}
\alpha^{(J)}_k = \langle x_n, g^{(J)}_{n-2^J k} \rangle_n = \sum_{n \in \mathbb{Z} } x_n ~g^{(J)}_{n-2^J k},	~~~~	J \in \mathbb{Z}
\label{eq:approximations}
\end{equation}

\begin{equation}
\beta^{(\ell)}_k = \langle x_n, h^{(\ell)}_{n-2^l k} \rangle_n = \sum_{n \in \mathbb{Z} } x_n ~h^{(\ell)}_{n-2^\ell k},	~~~~	\ell \in \{1, 2, ..., J\}
\label{eq:details}
\end{equation}

where $x_n$ is the $n^\textrm{th}$ input point, $\langle . \rangle$ is the dot product operation, and $g$'s and $h$'s are two sets of discrete orthogonal functions called the wavelet basis (we used the Haar basis functions in our system). 
 The inverse DWT is given by

\begin{equation}
x_n = \sum_{k \in \mathbb{Z} } \alpha^{(J)}_k ~g^{(J)}_{n-2^J k} + \sum^J_{\ell=1} \sum_{k \in \mathbb{Z} }\beta^{(\ell)}_k ~h^{(\ell)}_{n-2^\ell k}
\label{eq:idwt}
\end{equation}

\subsubsection{Noise Reduction}
Due to complex wireless propagation and interaction with surrounding objects, RSSI values at the receiver are noisy. This noise can add false edges and affect \sys{} accuracy and robustness. 
Therefore, to increase \sys{} quality, we leverage a wavelet-based denoising method \cite{sardy2001robust}. 
Wavelet denoising consists of three stages: decomposition, thresholding detail coefficients, and reconstruction.

In \textit{decomposition}, DWT is recursively applied to break the signal into high-frequency coefficients (details) and low-frequency coefficients (approximations) at different frequency levels.
\textit{Thresholding} is then applied to the wavelet detail coefficients to remove their noisy part. The threshold is chosen dynamically, based on minimizing the Stein unbiased risk estimate
(SURE), under the assumption of Gaussian noise \cite{sardy2001robust}. 
Finally, \textit{reconstruction} of the denoised signal occurs by combining the coefficients of the last approximation level with all thresholded details.

Wavelet denosing has the advantage of being computationally efficient (linear time complexity) to fit on mobile devices. In addition, it does not make any particular assumptions about the nature of the signal, and permits discontinuities in the signal \cite{wavlet_adv}. Moreover, we leverage DWT in other system functionalities, allowing code sharing for better efficiency.

\begin{figure}[!t]
\centering
        \begin{subfigure}[t]{0.23\textwidth}
                \centering
                \includegraphics[width=\textwidth]{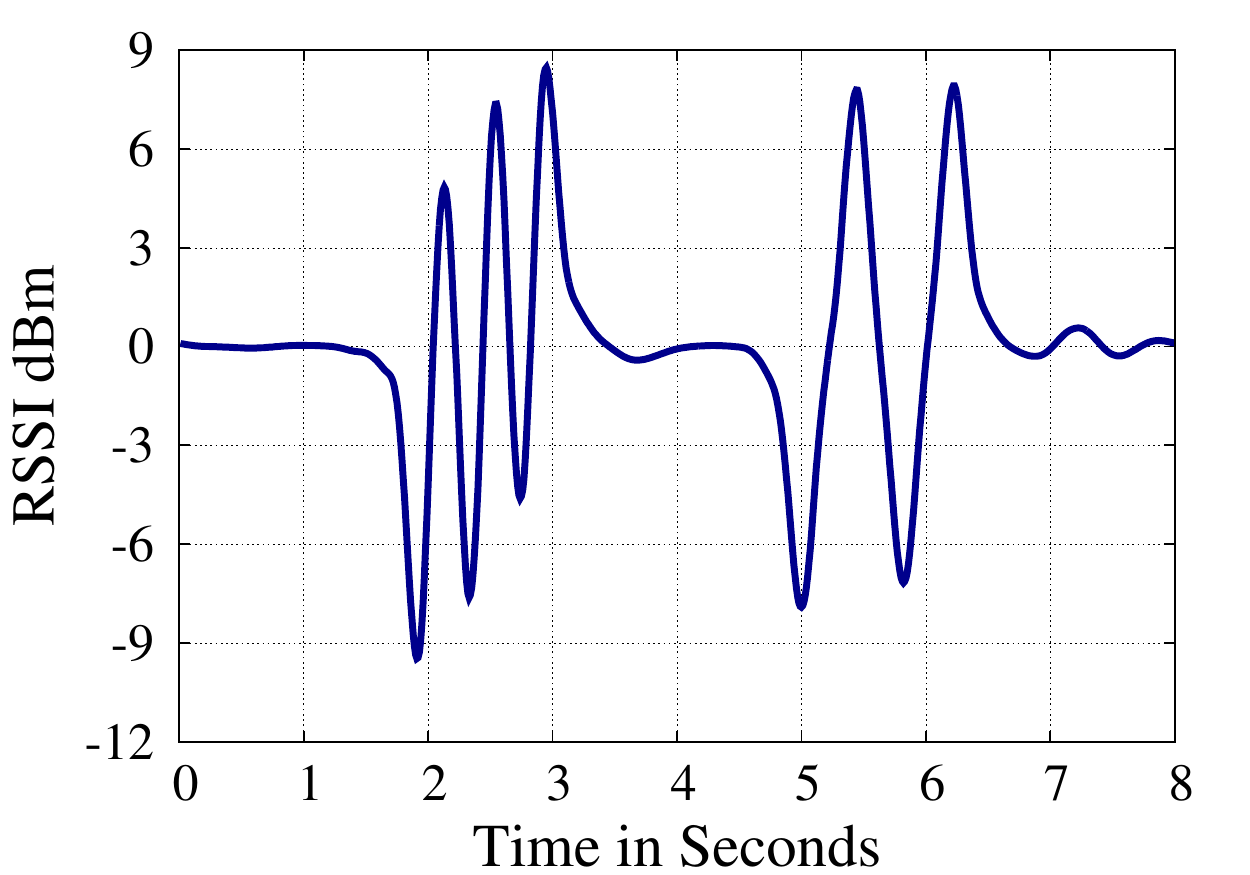}
                \caption{Detail coefficients of a denoised signal containing two motion speeds}
                \label{}
        \end{subfigure}
	    \begin{subfigure}[t]{0.23\textwidth}
                \centering
                \includegraphics[width=\textwidth]{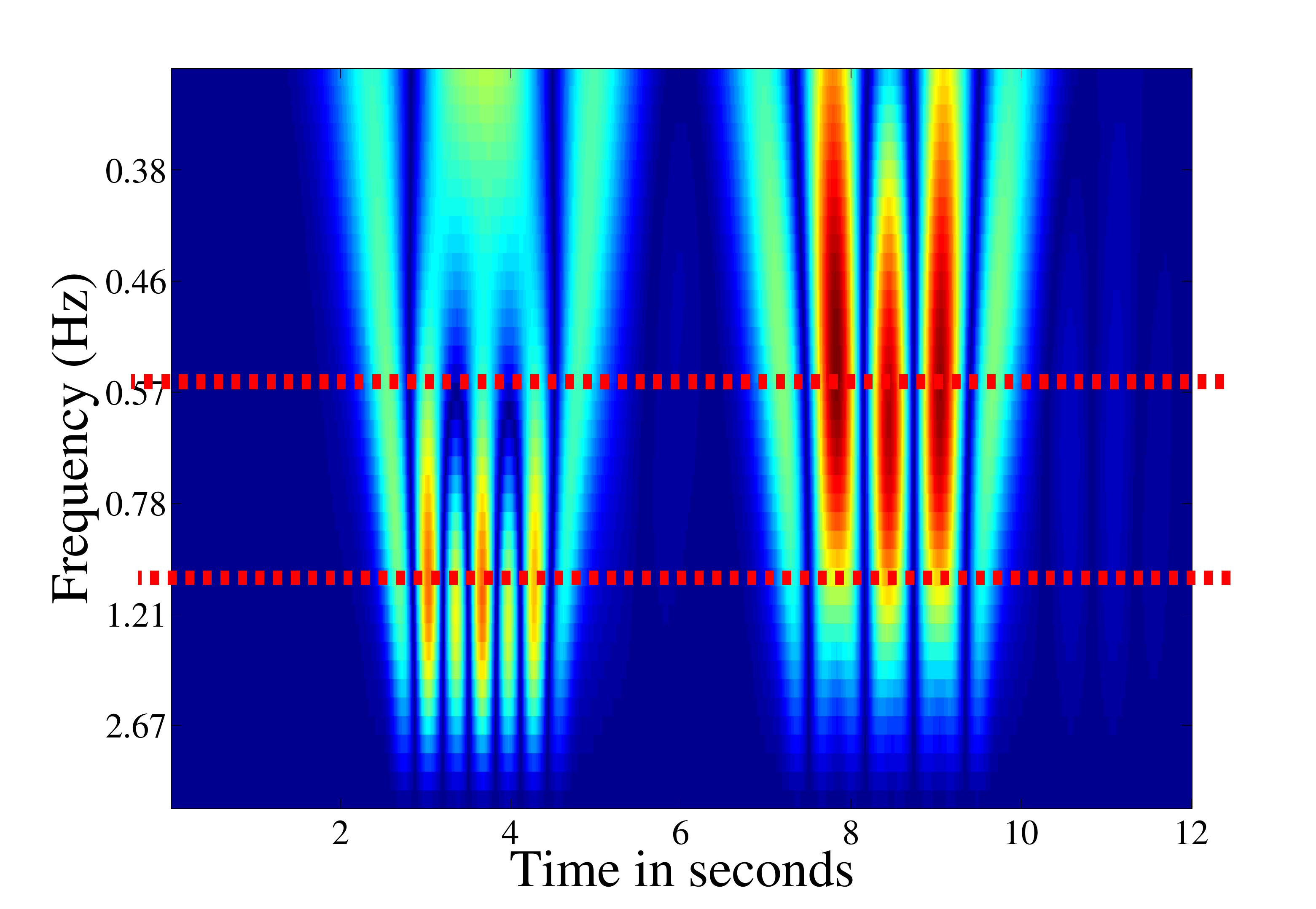}
                \caption{Spectrogram}
                \label{fig:spectrogram}
        \end{subfigure}
\caption{Detecting frequencies of interest dynamically using wavelet transform. The dashed lines in the spectrogram  map to the two frequencies of interest.}
\vspace{-0.2in}
\label{fig:freq_detection}
\end{figure}

\begin{figure*}[!t]
\centering
	\begin{subfigure}[b]{0.1\textwidth}
                \centering
                \includegraphics[width=\textwidth, height=5cm]{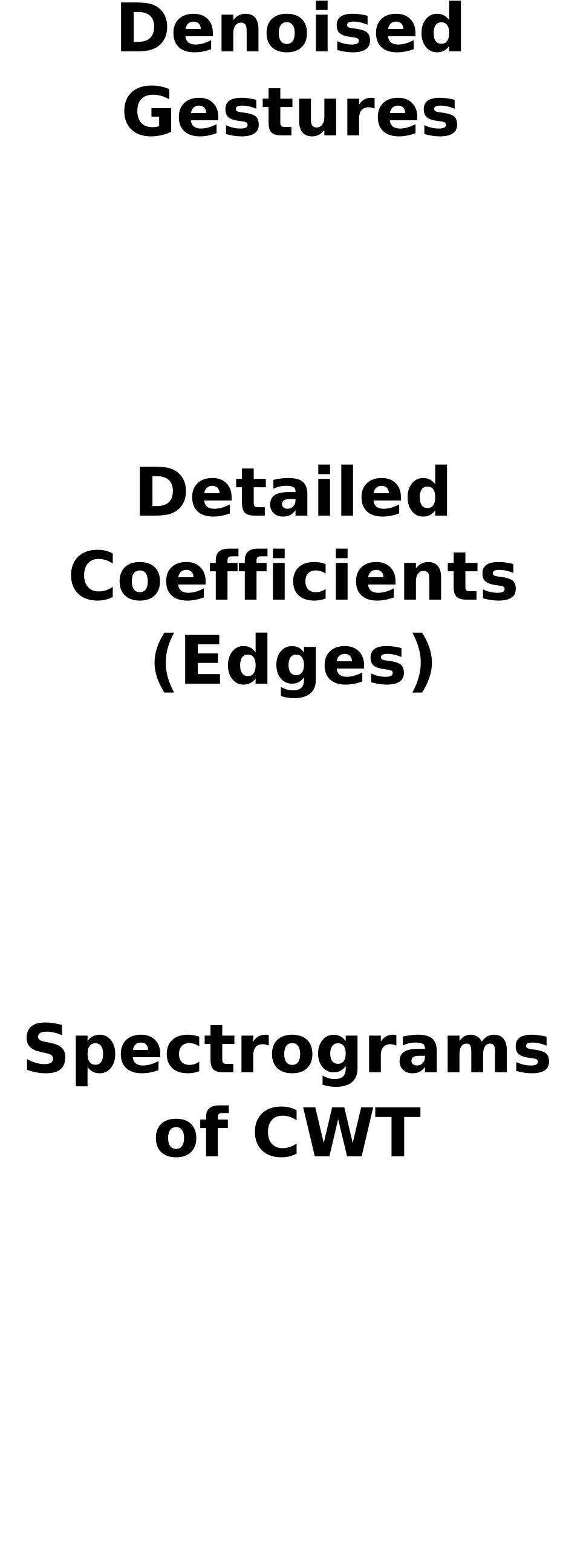}
        \end{subfigure}\hfill
        \begin{subfigure}[b]{0.123\textwidth}
                \centering
                \includegraphics[width=\textwidth, height=5cm]{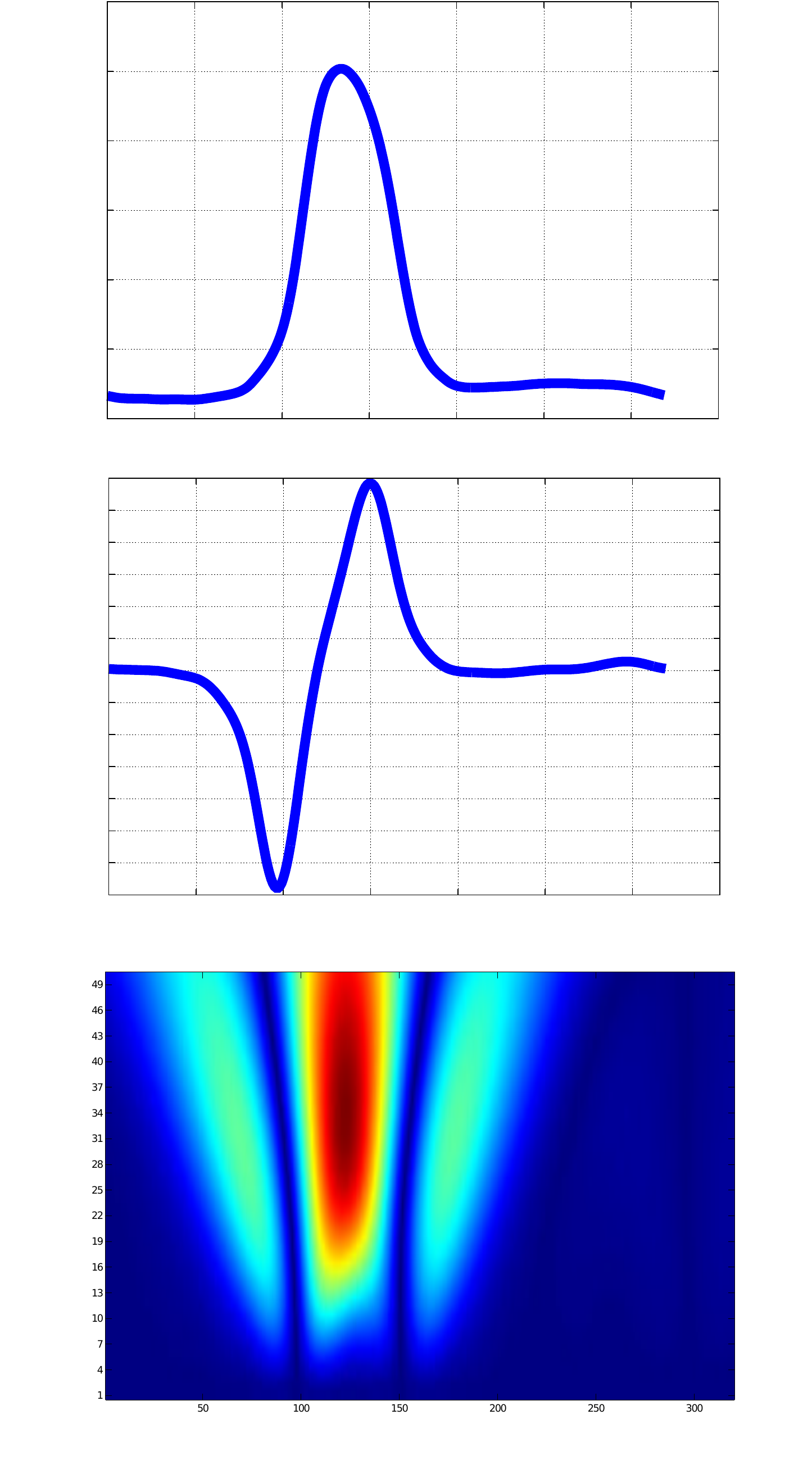}
                \caption{Up-Down}
                \label{fig:gestures_signs:ud}
        \end{subfigure}\hfill
        \begin{subfigure}[b]{0.123\textwidth}
                \centering
                \includegraphics[width=\textwidth, height=5cm]{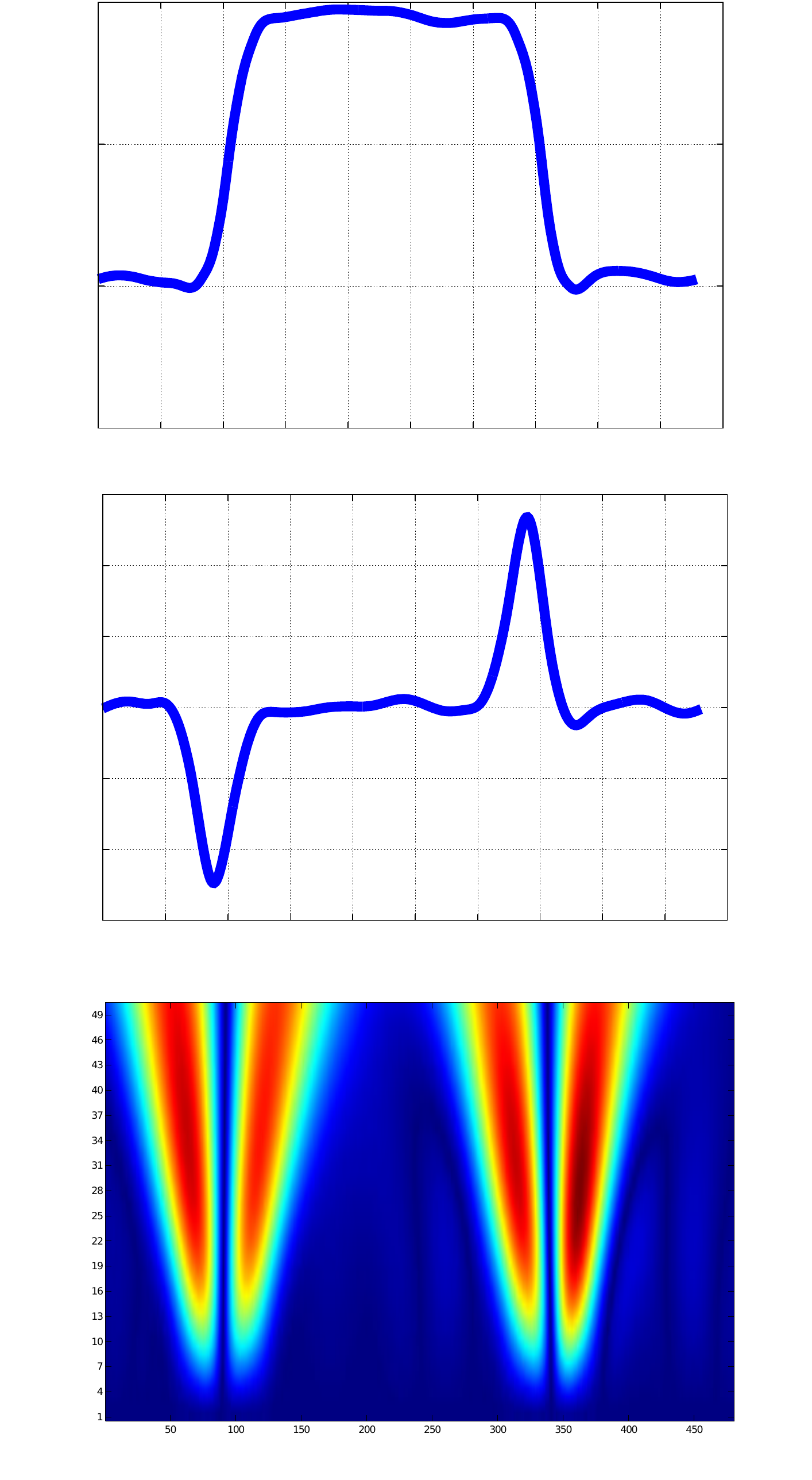}
                \caption{Up-$\|$-Down}
                \label{fig:gestures_signs:upd}
        \end{subfigure}\hfill
	\begin{subfigure}[b]{0.122\textwidth}
                \centering
                \includegraphics[width=\textwidth, height=5cm]{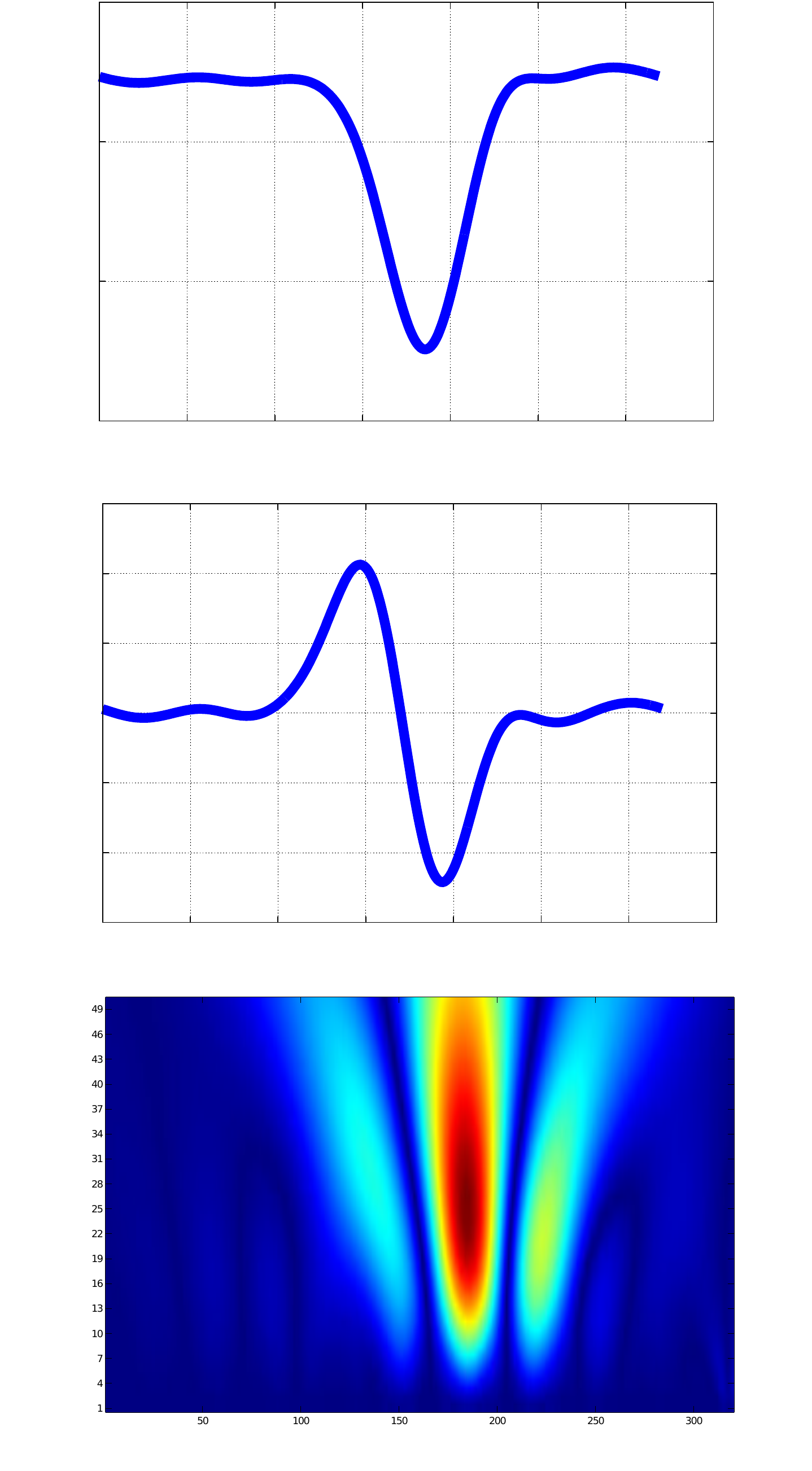}
                \caption{Down-Up}
                \label{fig:gestures_signs:du}
        \end{subfigure}\hfill
        \begin{subfigure}[b]{0.122\textwidth}
                \centering
                \includegraphics[width=\textwidth, height=5cm]{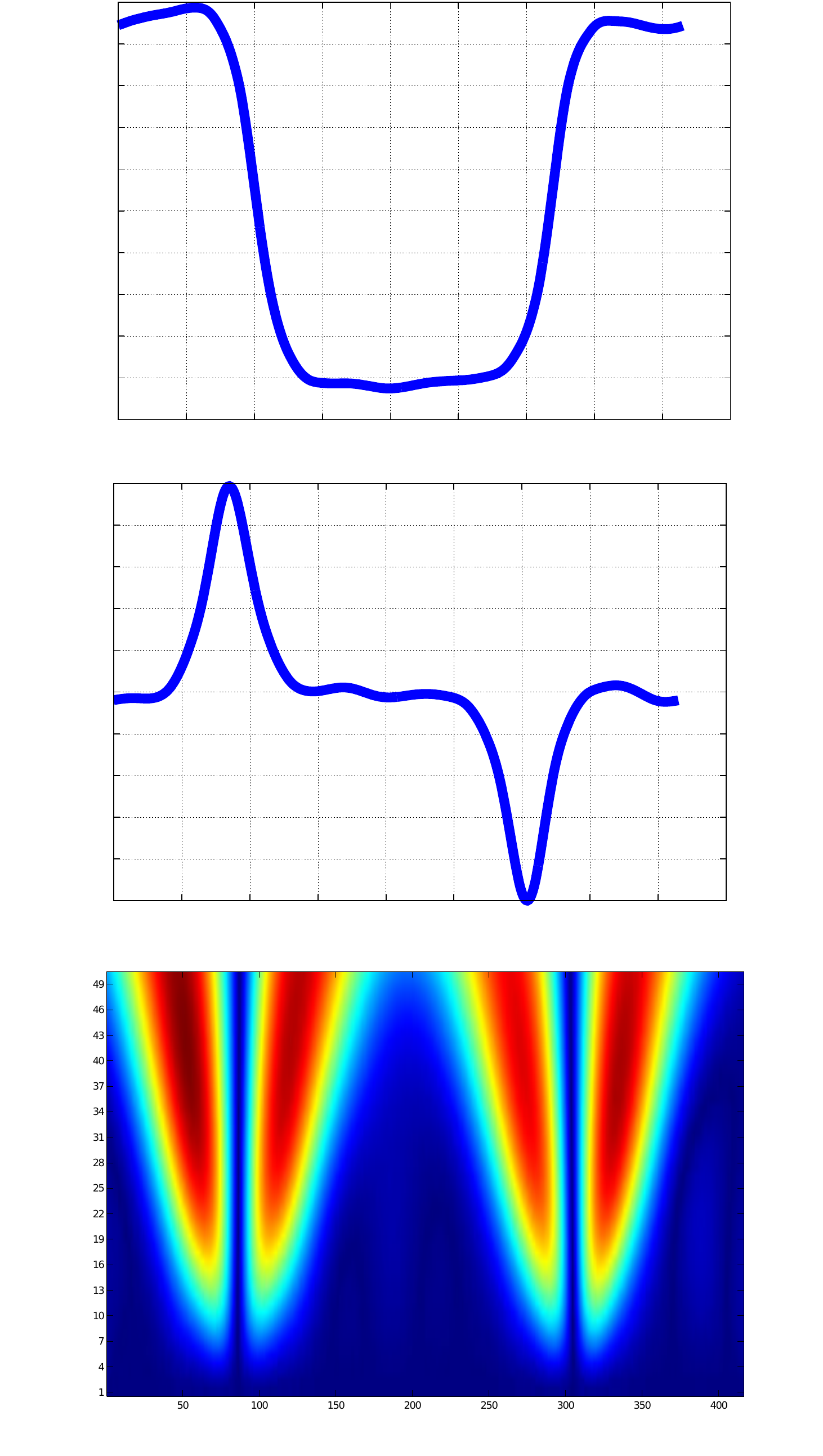}
                \caption{Down-$\|$-Up}
                \label{fig:gestures_signs:dpu}
        \end{subfigure}\hfill
        \begin{subfigure}[b]{0.122\textwidth}
                \centering
                \includegraphics[width=\textwidth, height=5cm]{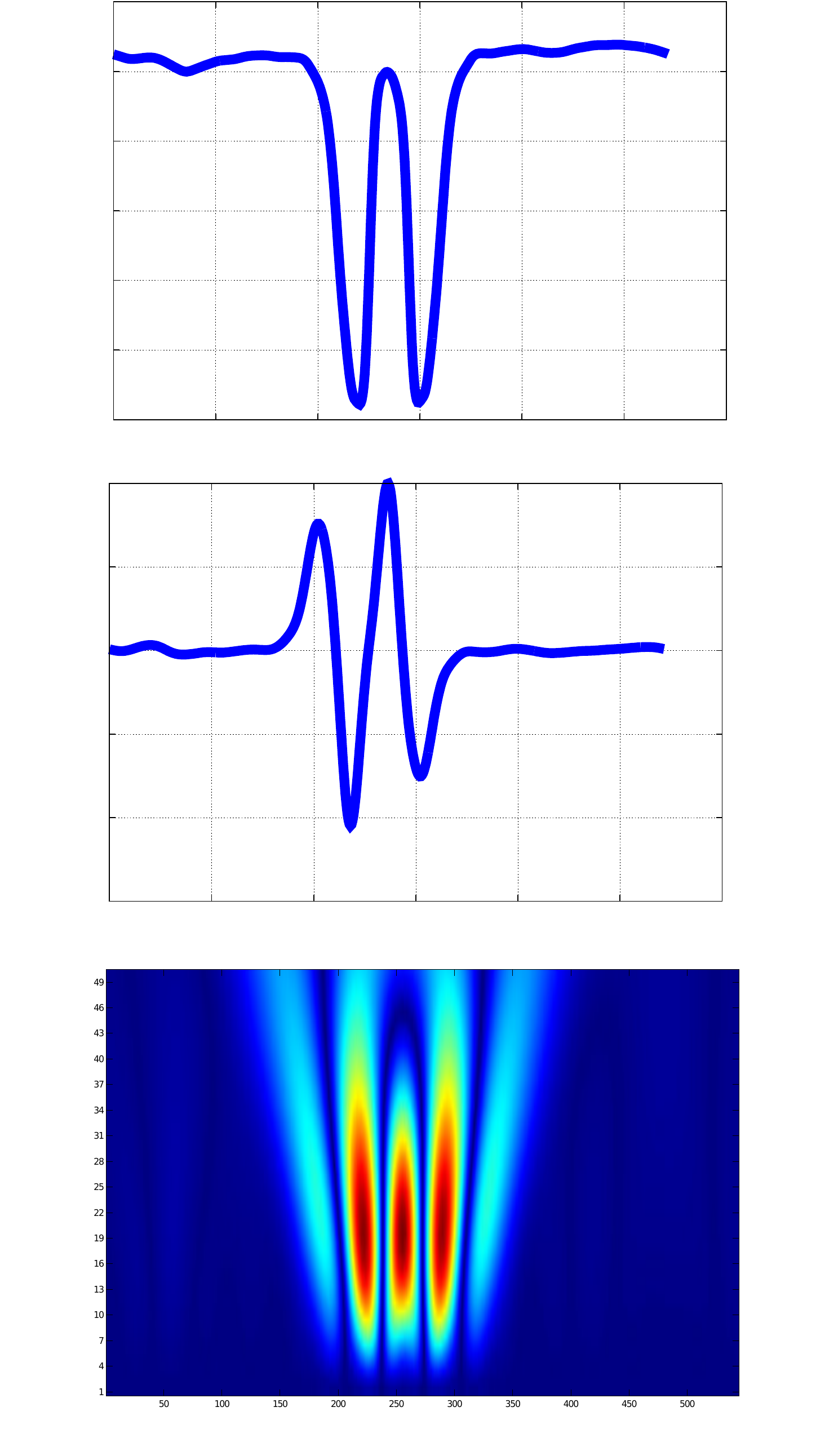}
                \caption{Right-Left}
                \label{fig:gestures_signs:rl}
        \end{subfigure}\hfill
        \begin{subfigure}[b]{0.123\textwidth}
                \centering
                \includegraphics[width=\textwidth, height=5cm]{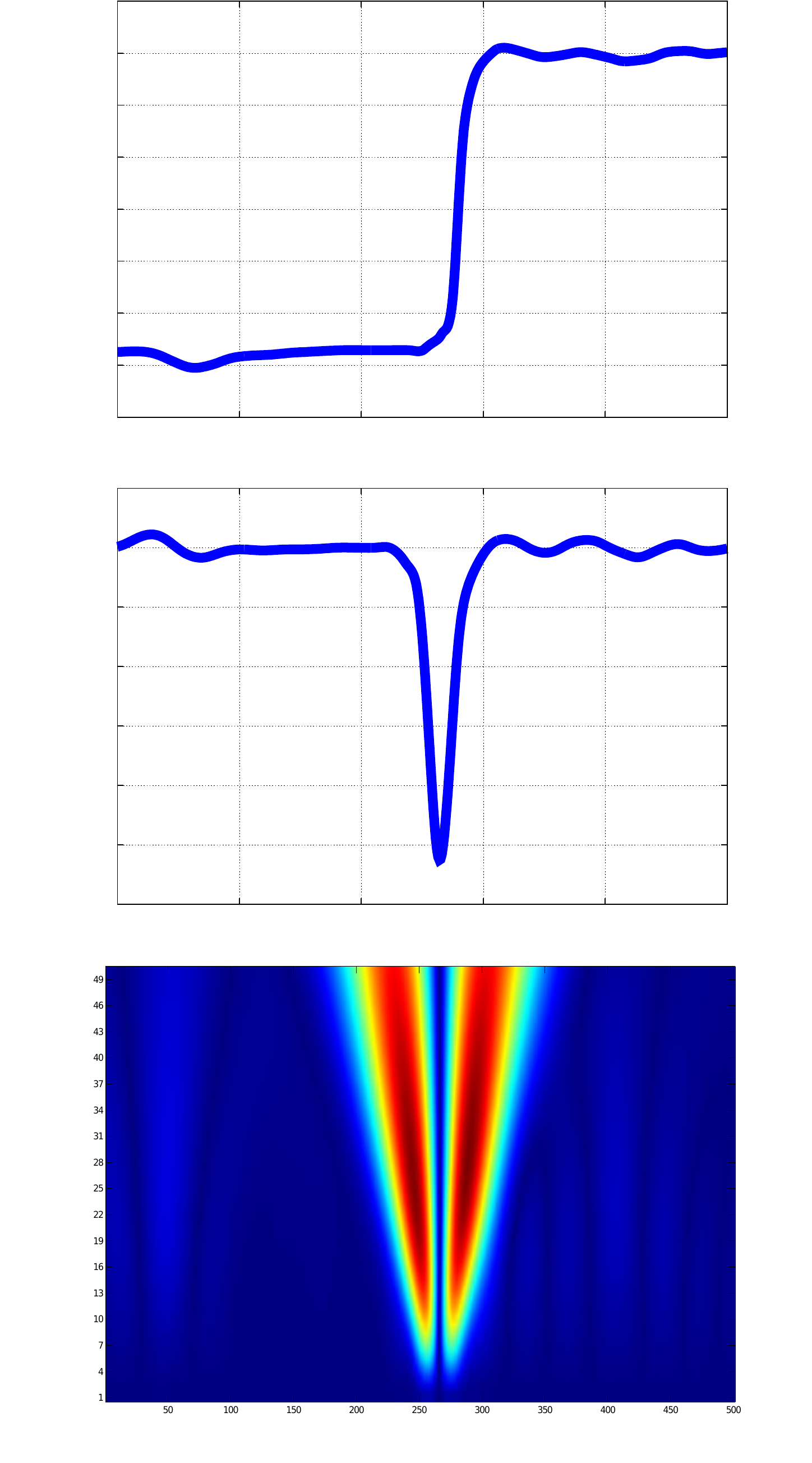}
                \caption{Up}
                \label{fig:gestures_signs:u}
        \end{subfigure}\hfill
        \begin{subfigure}[b]{0.122\textwidth}
                \centering
                \includegraphics[width=\textwidth, height=5cm]{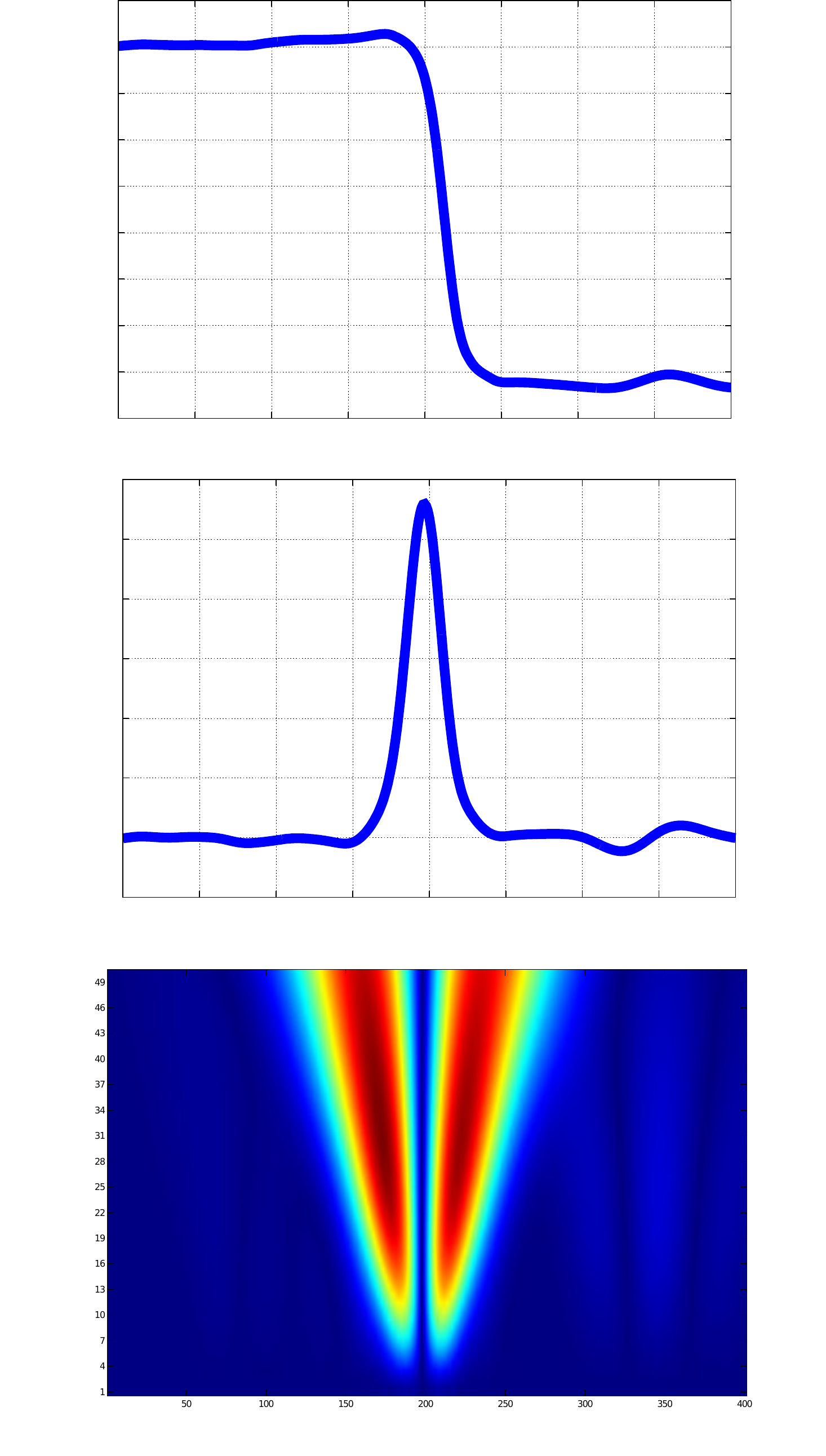}
                \caption{Down}
                \label{fig:gestures_signs:d}
        \end{subfigure}
	\caption{Different hand gestures, their detected edges, and spectrograms. The first row is the denoised RSSI signals. The second row shows the fifth level details coefficients of the denoising module used in edge detection. The third row shows the spectrogram of the signal only used to determine the frequency of interest level (not for distinguishing gestures.)}
\vspace{-0.2in}
\label{fig:gestures_unites}
\end{figure*}

\subsubsection{Edge Extraction and Primitives Detection}
The main effect of the human hand motions on the received signal (detection primitives) are either rising edges, falling edges, or pauses. The place of these primitives need to be defined in time.
However, since the human hand motion speed may be different from one person to another or for the same person at different times, we need a technique that can provide variable frequency resolution. Therefore, we use wavelet analysis for edge detection in \sys{}. Compared to other techniques, e.g. the Fourier transform, Wavelet analysis provides both time and frequency locality at different resolutions as well as being computationally efficient.

Figure~\ref{fig:freq_detection} shows the spectrogram of a denoised signal containing two different hand motion velocities. The figure shows that Wavelet transform can capture/localize the two frequencies of motion (the two dashed lines in the figure) using simple 2D local maximum extraction technique. This technique is then utilized in the DWT process to determine the levels of interest that are used in the edge detection stage.

Based on the DWT theory, a rising edge causes a local minima peak and a falling edge causes a local maxima peak in the detailed coefficients as shown in the second row of Figure~\ref{fig:gestures_unites}. A pause is detected as a stable RSSI value (RSSI variance within a threshold) for a certain minimum time (0.5 second in our system). Therefore, using this module, any input signal can be translated into a set of rising edges, falling edges, and/or pauses. Figure~\ref{fig:gestures_unites} shows some examples of different hand motion gestures and the detected edge positions as well as their spectrograms. The figure also shows that the edge locations and frequency can be accurately extracted.

We note that \sys{} can use multiple overheard APs to increase the system accuracy and resilience to noise. In this case, a majority vote on the detected primitive is used to fuse the detection of the different APs.

\subsubsection{Primitives attributes}
After extracting edges, each edge can be labeled with two parameters: speed and magnitude. Based on the edge duration, there are three possible speeds: high (edge duration $< 0.75$ sec), medium ($0.75$ sec $<$ edge duration  $<1.5$ sec), or low (edge duration $>1.5$ sec). On the other hand, the distance value is also discretized into two values, high (more than 0.55 $\textrm{ft}$ above the receiver) and low (less than 0.55 $\textrm{ft}$ above the receiver), which is determined based on a threshold applied to the RSSI. Since the amplitude of the signal change is relative to the original signal level (i.e. a strong signal will lead to a lager change with the user hand movement and vice versa), we base the threshold value on the change of the start of the signal as we describe in the signal segmentation section below.
\subsection{Gesture Identification}
The goal of this processing stage is to extract the different gestures and their attributes (frequency and count). This is performed through two steps: segmentation and identification/matching.

\subsubsection{Segmentation}
One important question for the correct operation of \sys{} is when to know that the user is generating a gesture. This helps avoid false gestures generated by other actions/noise in the environment from nearby users and also leads to energy-efficiency. \sys{} answers this question by using a special preamble that is hard to confuse with other actions in the environment. Without this preamble, the system refrains from interpreting gestures. The preamble contains two states: a drop in the input signal values and then two up-down signals. This is equivalent to holding the hand over the receiver and then making two up-down gestures. The first stage is detected by a simple and efficient thresholding operation. So in real-time, the default case is that the preamble extraction module only searches for a drop in the RSSI values. If detected, it moves to the second stage and searches for two consecutive up-down motions, which is equivalent to searching for four consecutive peaks in the detailed components of the DWT (computed in linear time). Once the preamble is detected, the communication channel begins between the device and the user, and the system scans for various gestures based on the primitives extracted in the previous section. Gesture recognition is terminated after a silence period preset by the user, returning to the preamble scanning phase.

In some cases, due to the complex wireless propagation environment, flipping between rising and falling edges may occur. To compensate for this, \sys{} leverages the preamble. Specifically, the direction of the change of the first state (typically expected to be a drop in signal strength) determines whether the signal should be flipped or not. \sys{} also leverages the preamble to determine the threshold magnitude (high/low) and motion frequency. Essentially, the extracted features from the predefined preamble help the system adapt to different users and different environments.

\subsubsection{Pattern Encoding and Matching/Gesture Identification}
\label{sec:pattern_enc}
Once the gesture boundary is determined and primitives are extracted, we convert the primitives to a string sequence: rising edges to positive signs, falling edges to negative signs, and pauses to zeros. The extracted string pattern is then compared with gesture templates to find the best match, as well as extract the count and frequency attributes (i.e. number of gesture repetitions per unit time). 

\subsection{Action Mapping}
This is a direct mapping step based on the application semantics. The developer determines which application actions are mapped to which gesture families. In addition, the attributes of the gesture are passed to the application for further processing. For example, the frequency attribute can be used to determine how fast the character should run in a game, the count parameter can determine how far we should go in the list of choices, etc. Note that multiple actions can be mapped to the same gesture family if they have multiple attributes. For example, in a media player application, ``play'' can be mapped to a right-hand gesture, while fast forward can be mapped to a double right-hand gesture. 

\begin{figure}[!t]
\centering
	\begin{subfigure}[b]{0.3\textwidth}
	\includegraphics[width=\textwidth]{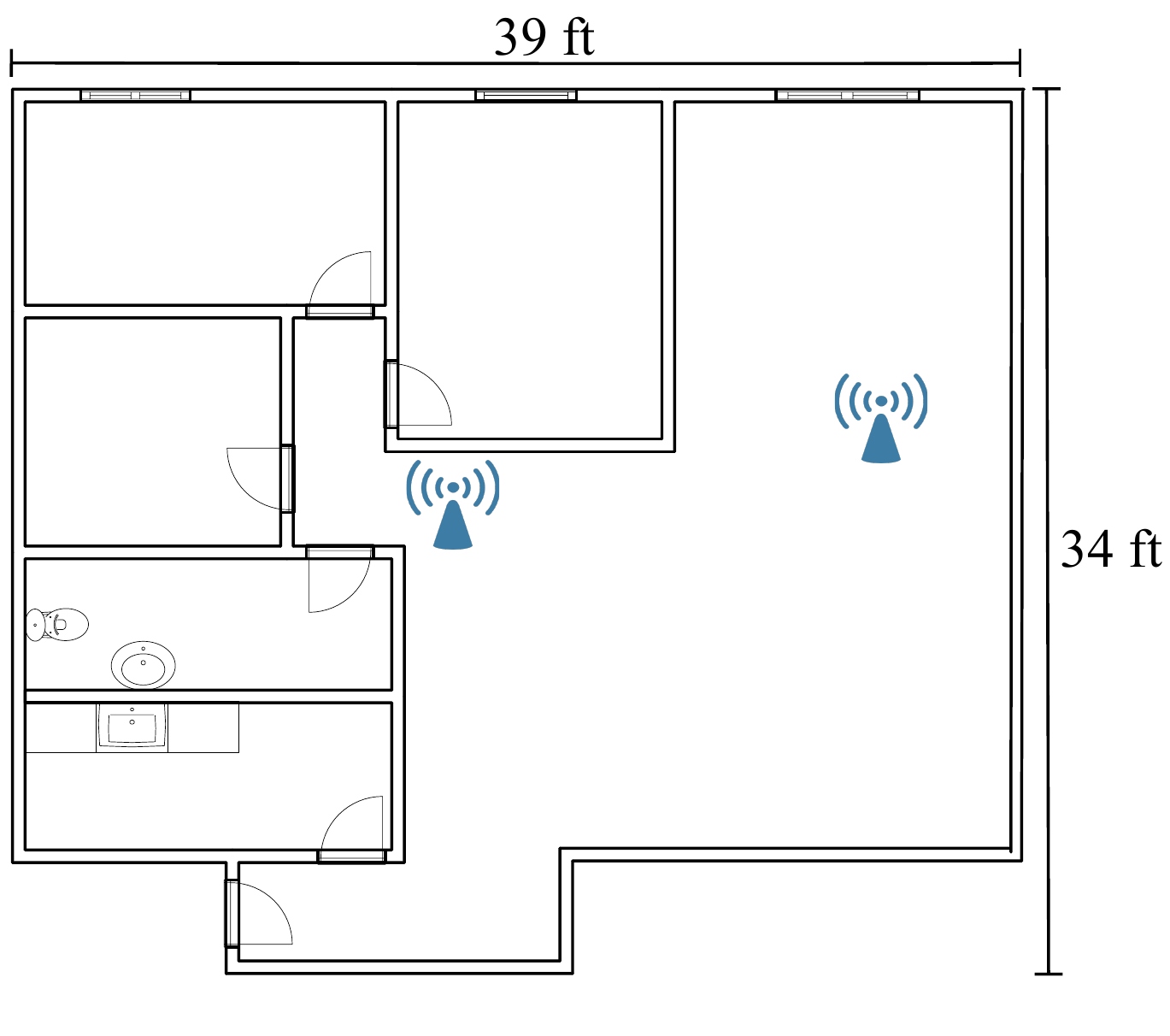}
	\caption{First test environment (apartment).}
	\label{fig:apartment_floorplan}
	\end{subfigure}
\centering
	\begin{subfigure}[b]{0.35\textwidth}
	\includegraphics[width=\textwidth]{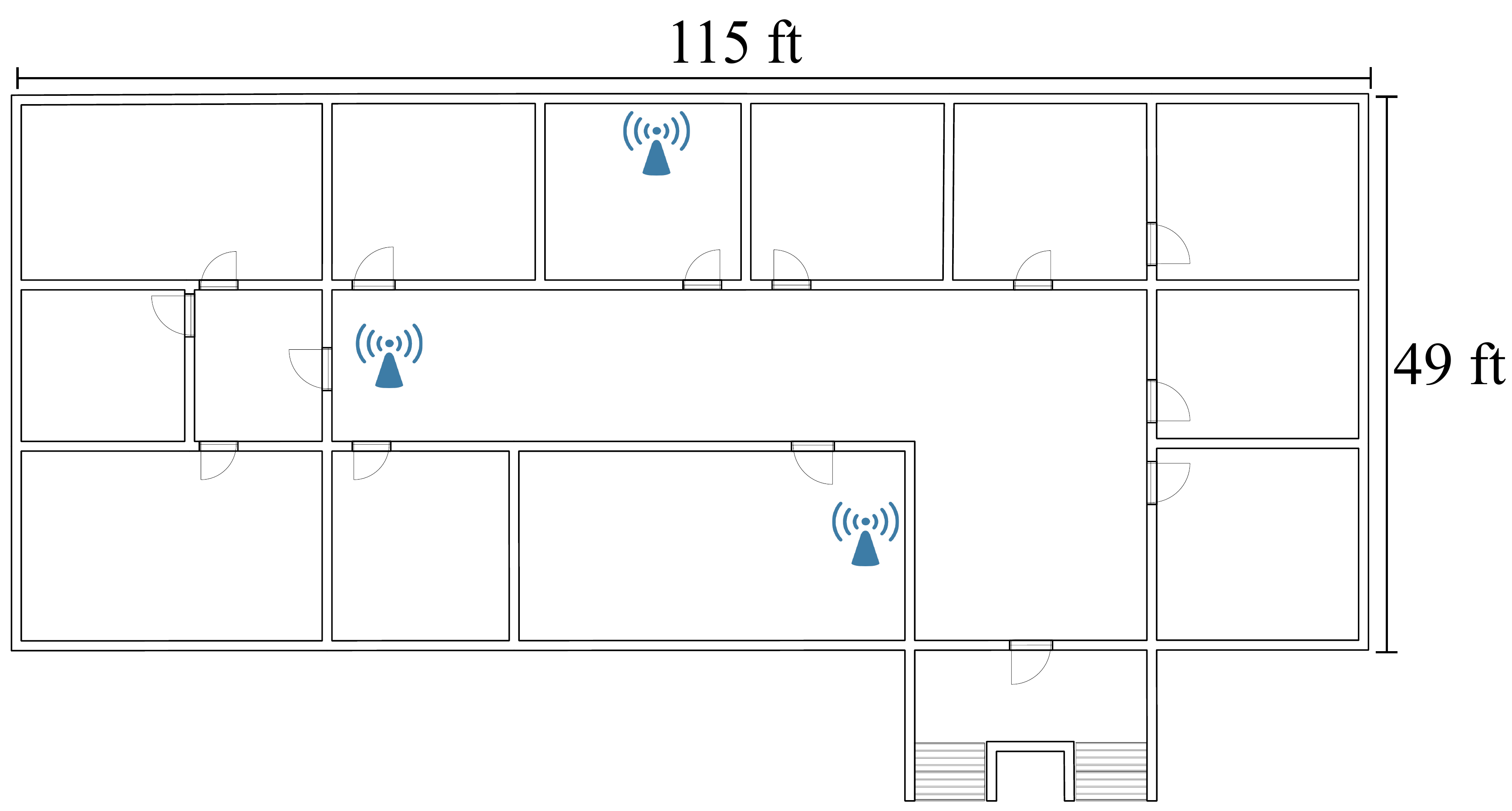}
	\caption{Second test environment (one floor of the engineering building where our lab is located). 
The APs shown are the ones installed in this floor.}
	\label{fig:vtmena_floorplan}
	\end{subfigure}
\caption{Floorplans of the two test environments used for evaluating \sys{}.}
\label{fig:testbeds}
\end{figure}

\begin{figure}[!t]
\centering
\includegraphics[width=3in]{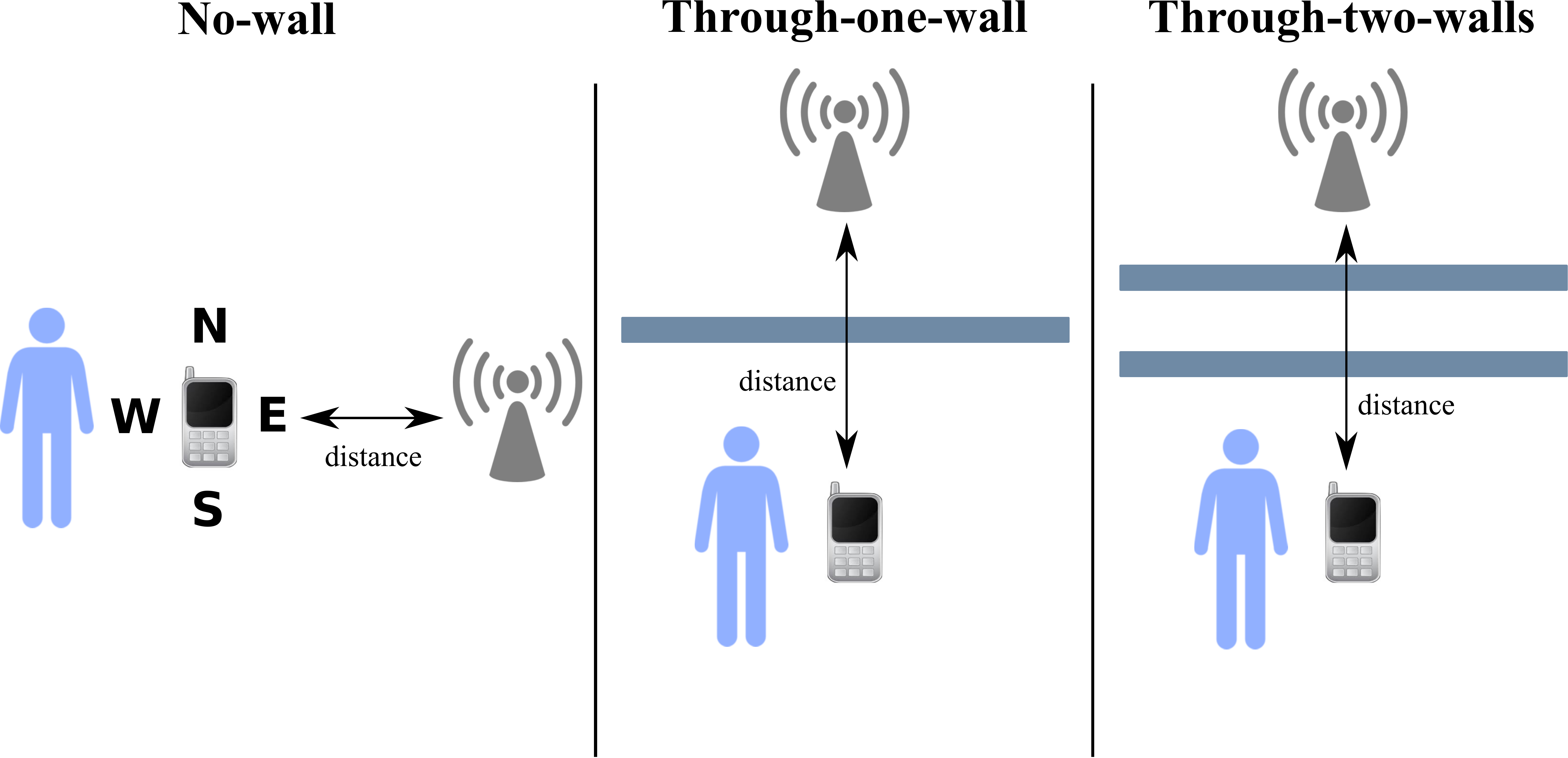}
\caption{The three scenarios layouts: the orientations characters in no-wall scenario indicate the possible four user orientations around her device relative to the AP.}
\vspace{-0.2in}
\label{fig:scenarios}
\end{figure}

\begin{figure}[!t]
\centering
\includegraphics[width=0.66\linewidth]{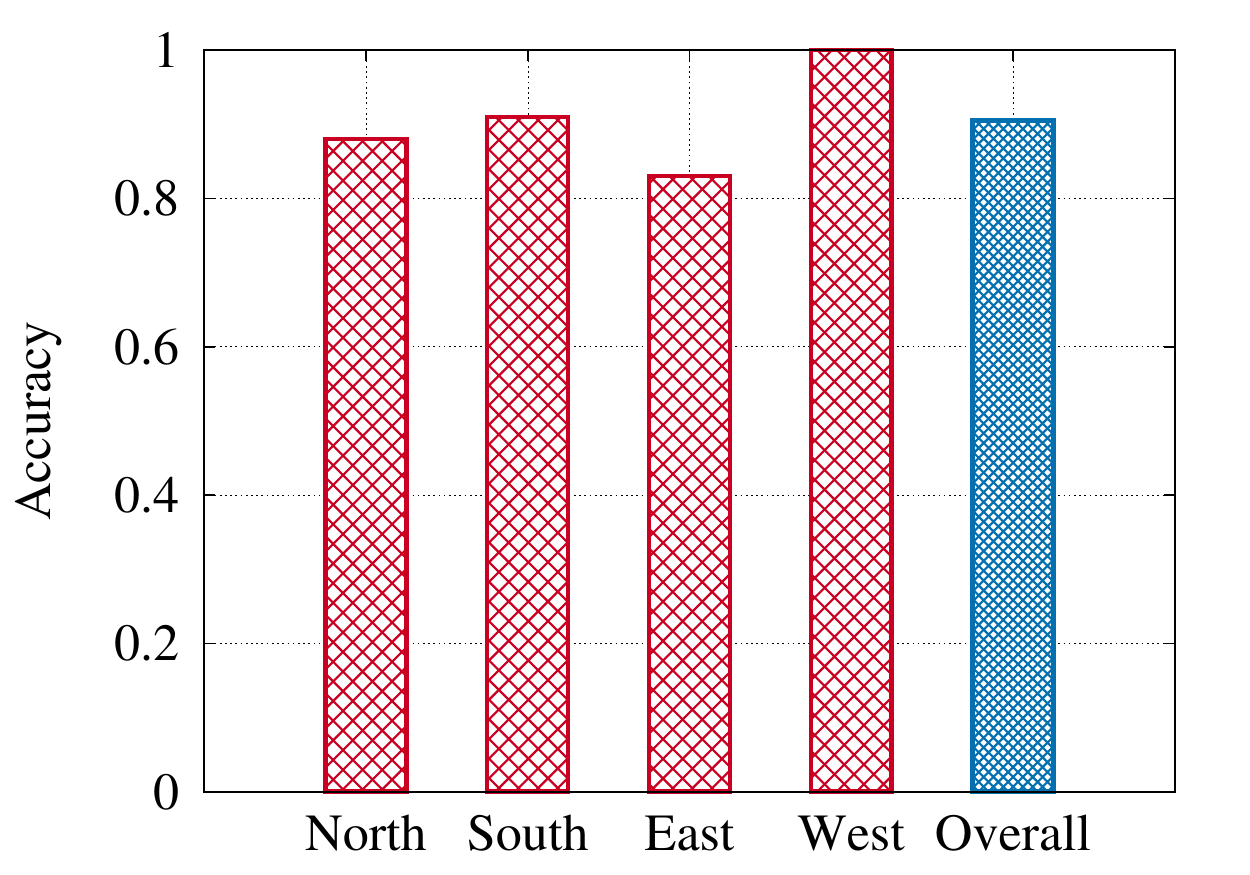}
\caption{Effect of the user location relative to the LOS between the device and the transmitter. The orientations are indicated in Figure \ref{fig:scenarios}.}
\label{fig:orientations}
\end{figure}

\begin{figure}[!t]
  \centering
  \includegraphics[width=0.66\linewidth]{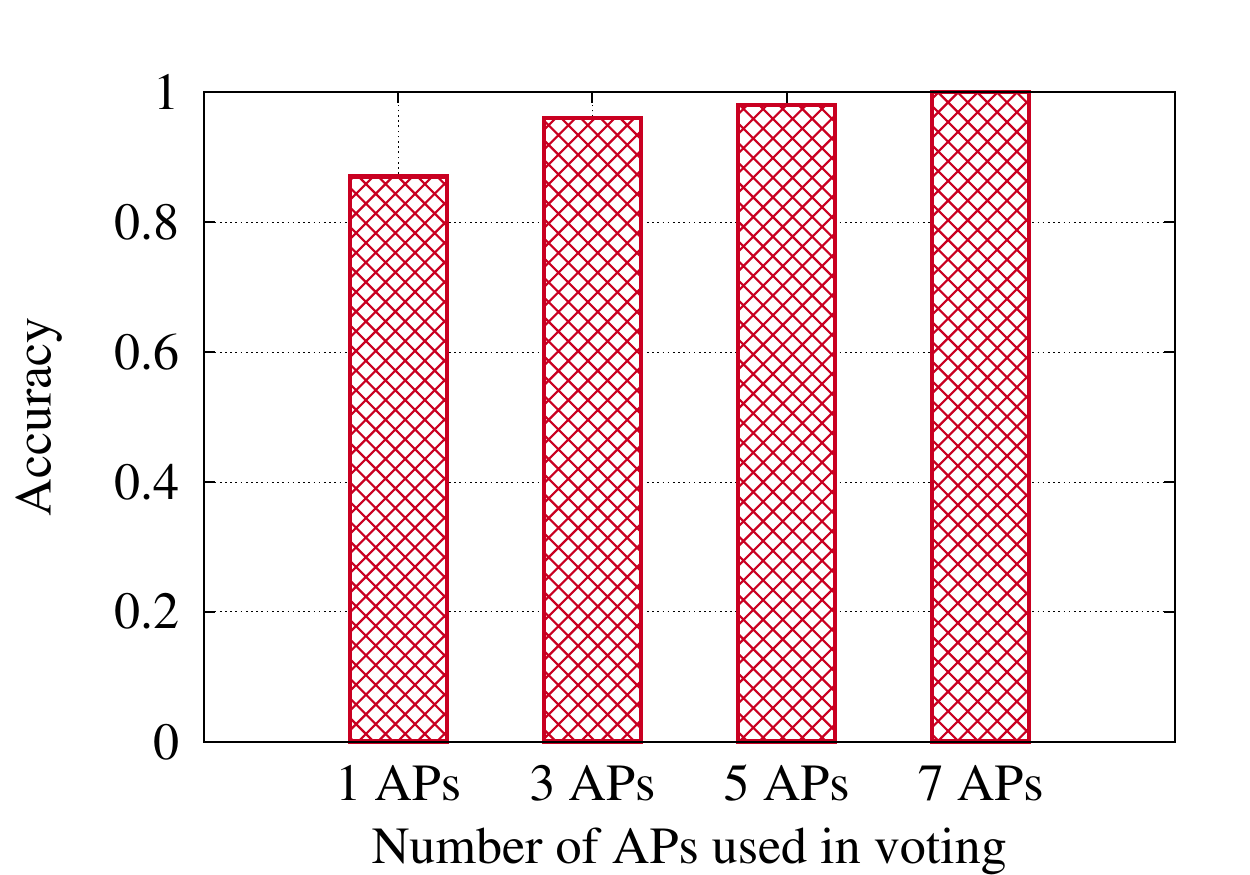}\\
  \caption{Effect of increasing the number of APs on the system accuracy.}\label{fig:multiple_aps}
\end{figure}

\section{Evaluation}
\label{evaluation}

\begin{figure*}[!t]
\centering
	\begin{subfigure}[b]{0.33\textwidth}
	\centering
	\includegraphics[width=\textwidth]{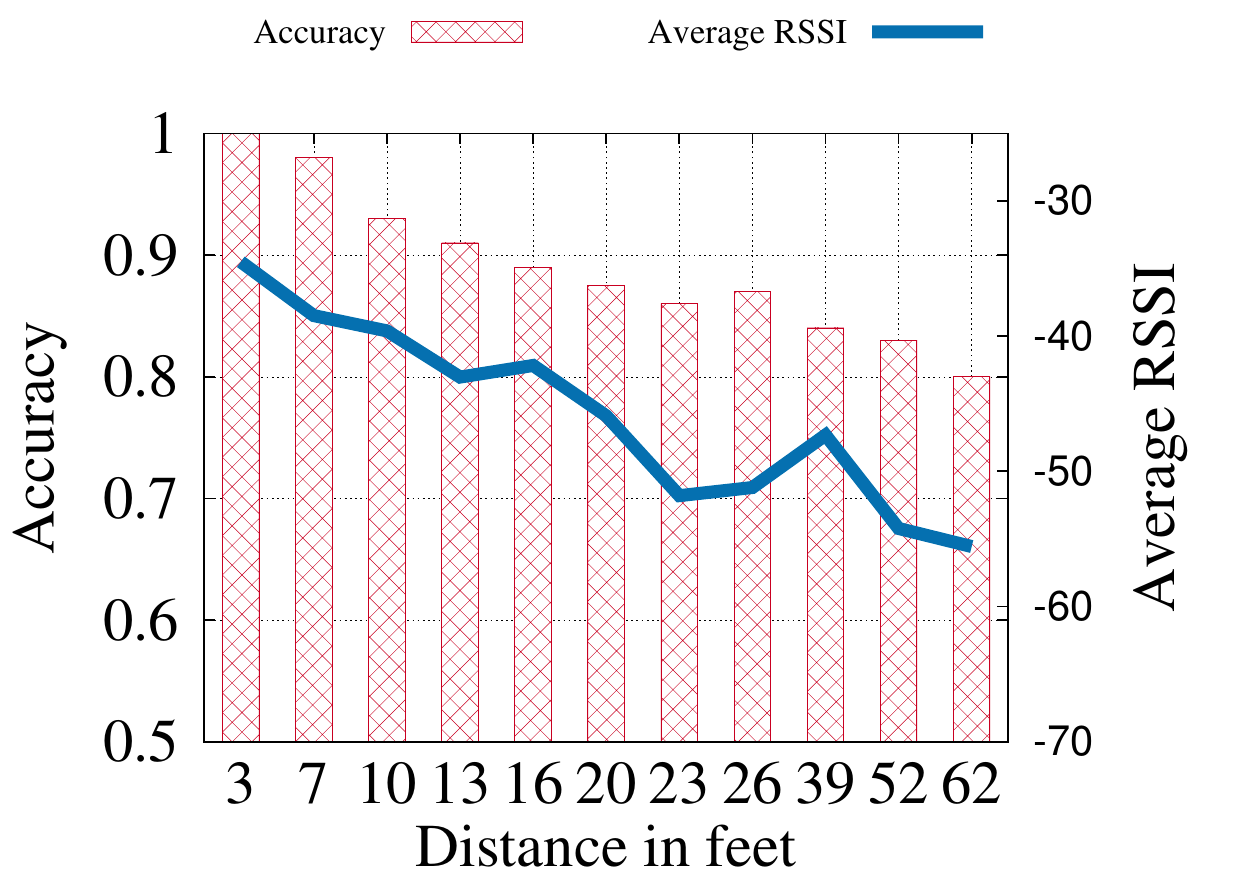}
	\caption{No-wall}
	\label{fig:dgd:no-wall}
	\end{subfigure}~
\centering
	\begin{subfigure}[b]{0.33\textwidth}
	\centering
	\includegraphics[width=\textwidth]{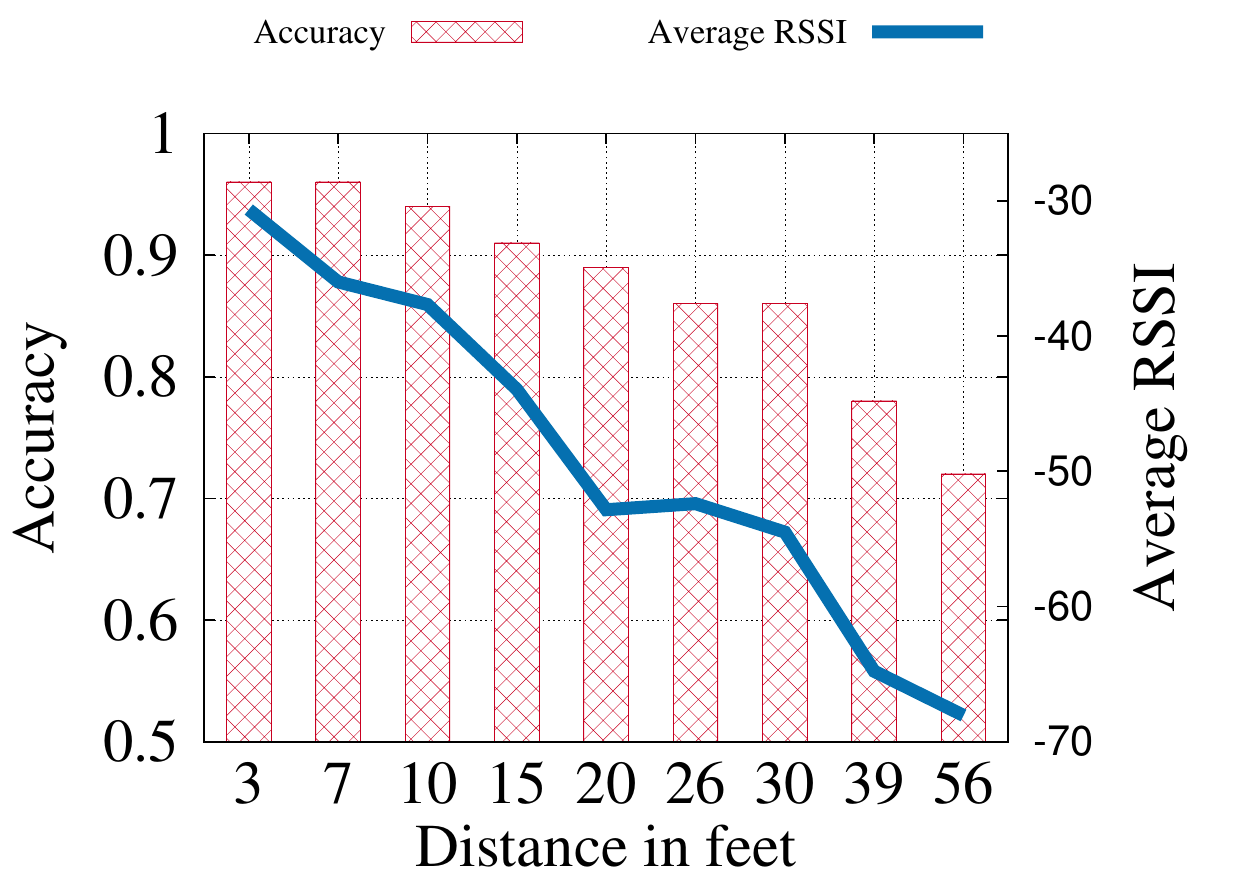}
	\caption{Through-one-wall}
	\label{fig:dgd:through-one-wall}
	\end{subfigure}~
\centering
	\begin{subfigure}[b]{0.33\textwidth}
	\centering
	\includegraphics[width=\textwidth]{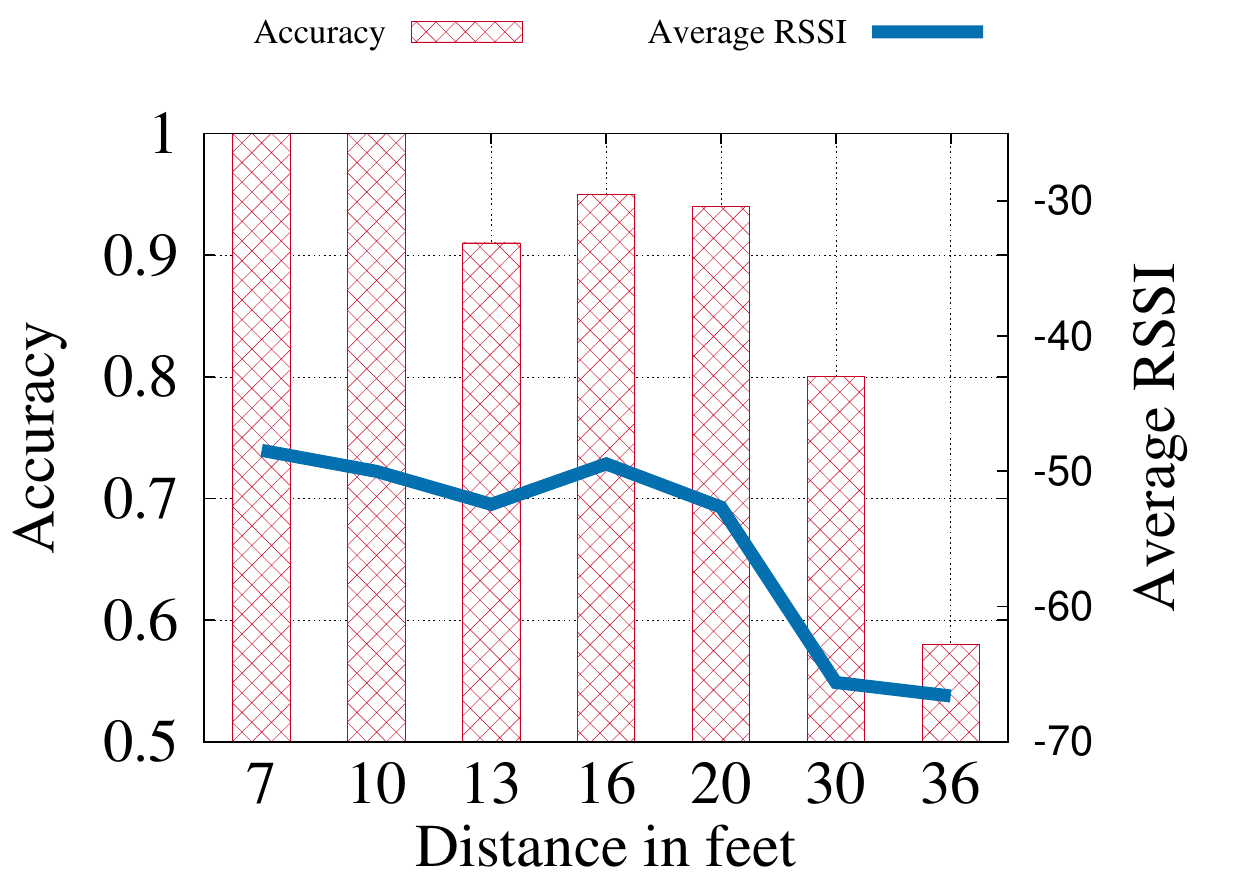}
	\caption{Through-two-walls}
	\label{fig:dgd:through-two-walls}
	\end{subfigure}

\caption{Impact of distance on edge detection accuracy. The figure also shows the average RSSI value at each distance.}
\label{fig:distance_gesture_detection}
\end{figure*}

\begin{figure*}[!t]
\centering
	\begin{subfigure}[b]{0.3\textwidth}
	\centering
	\includegraphics[width=\textwidth]{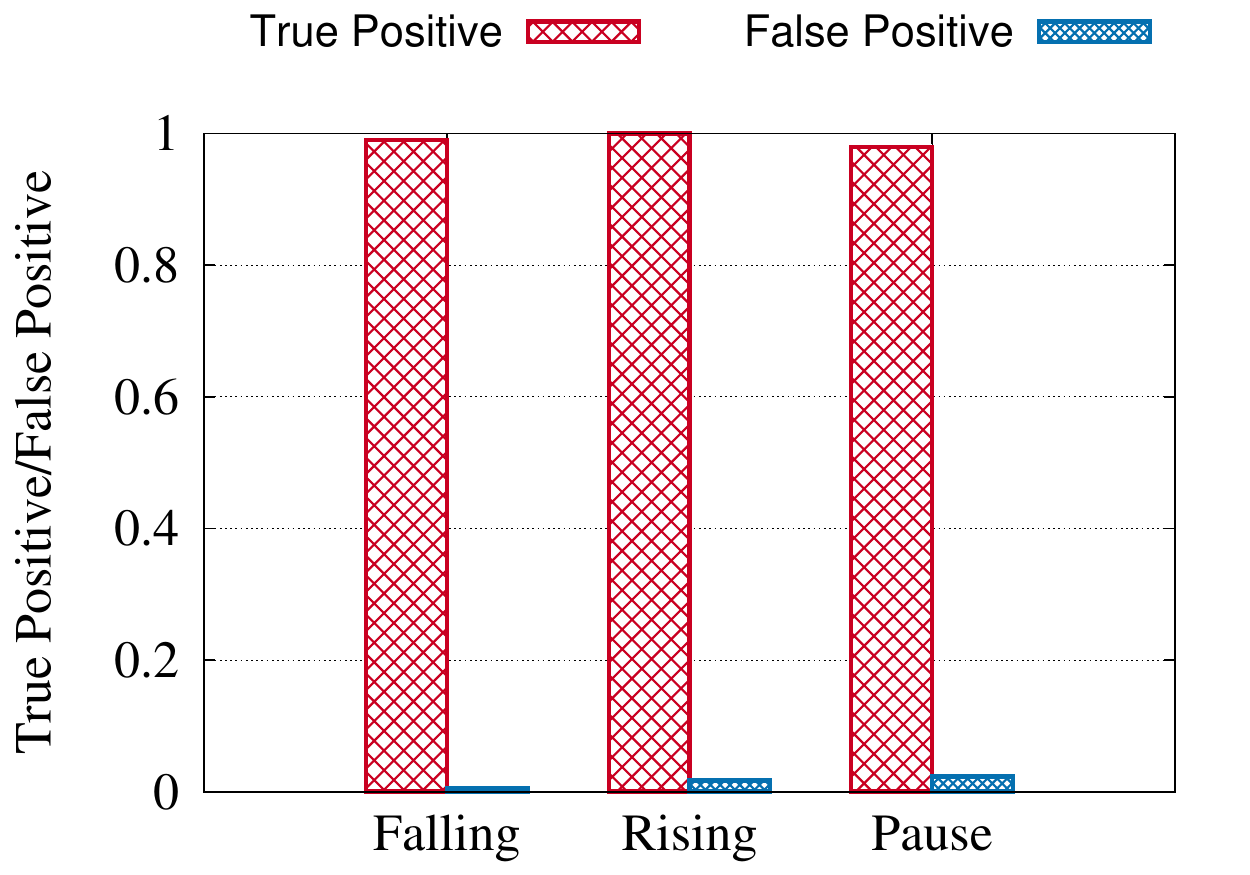}
	\caption{Primitives}
	\label{fig:prim:distinction}
	\end{subfigure}~
\centering
	\begin{subfigure}[b]{0.3\textwidth}
	\centering
	\includegraphics[width=\textwidth]{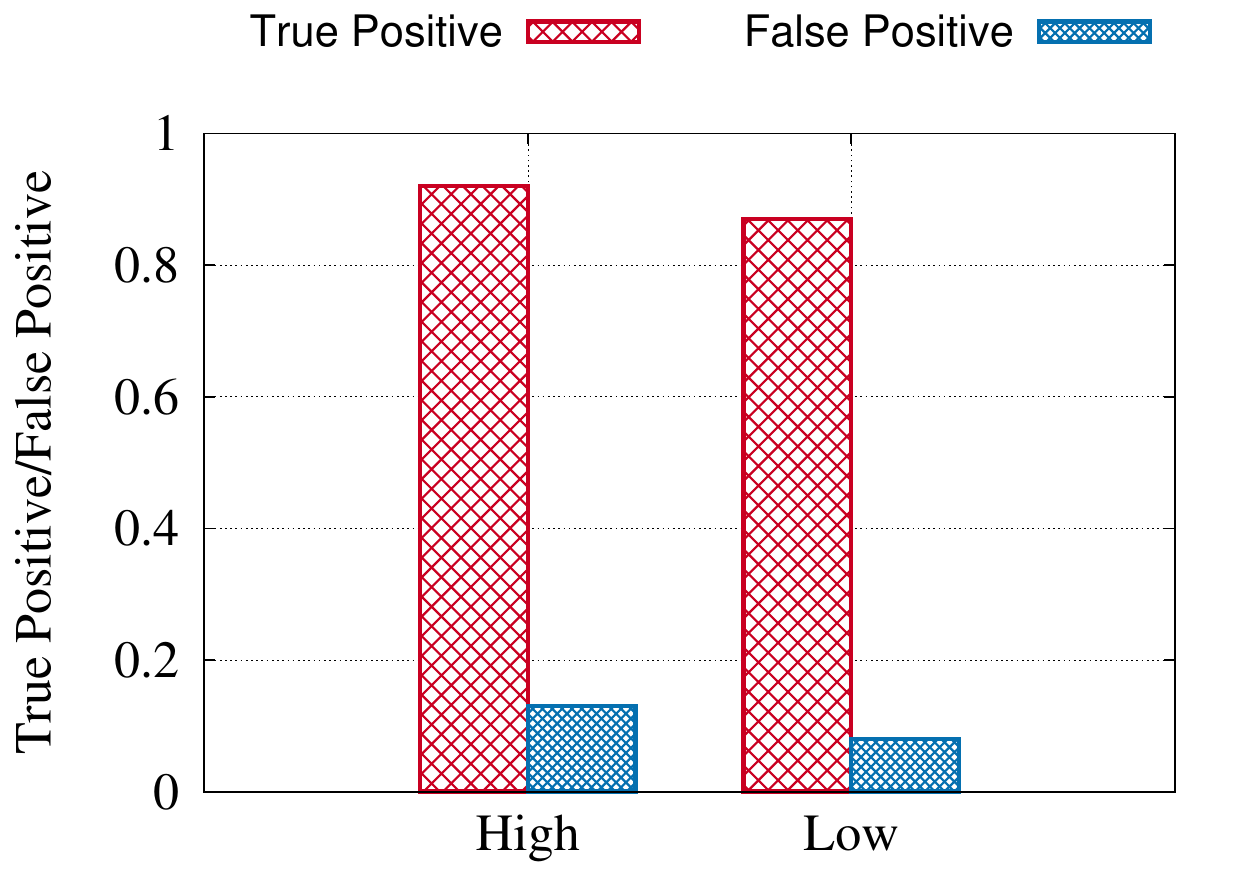}
	\caption{Edge magnitude}
	\label{fig:prim:magnitude}
	\end{subfigure}~
\centering
	\begin{subfigure}[b]{0.3\textwidth}
	\centering
	\includegraphics[width=\textwidth]{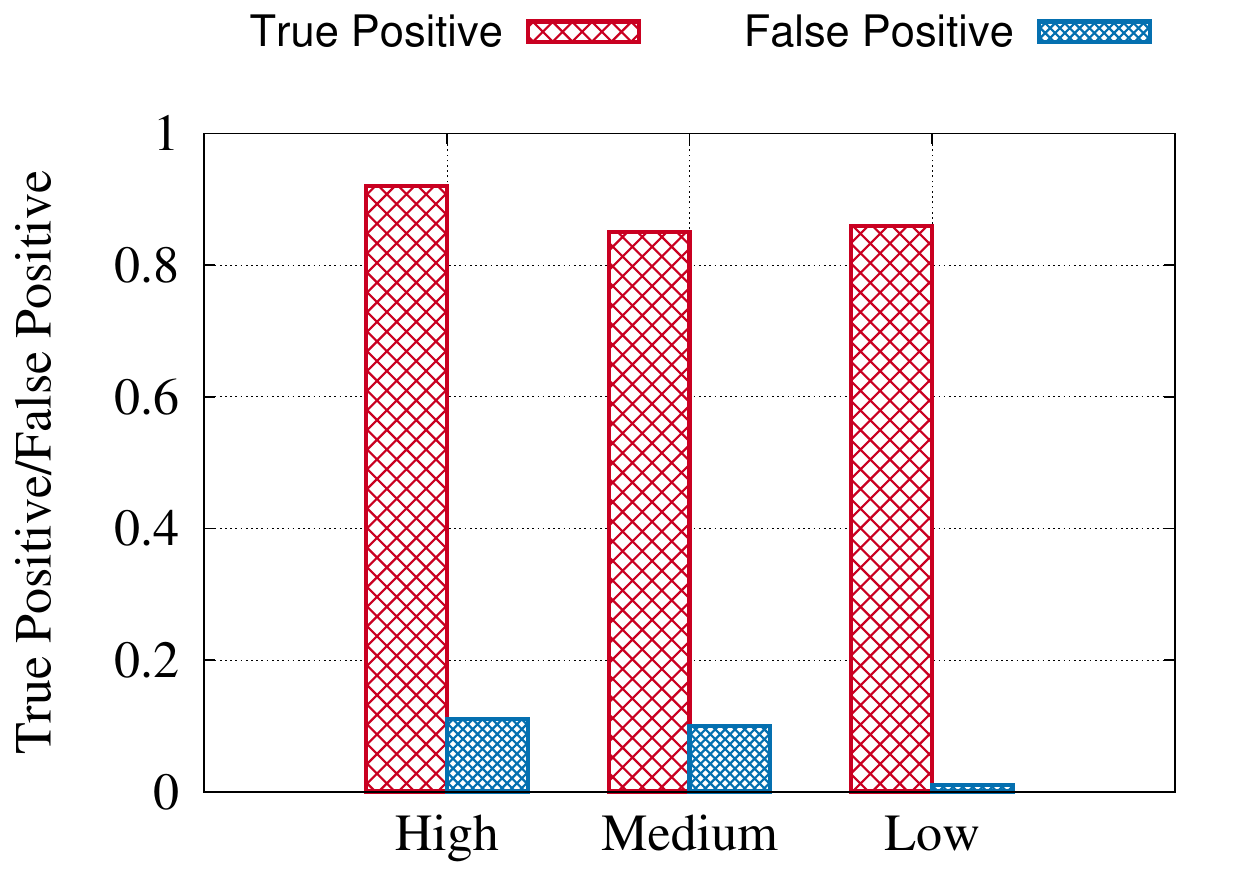}
	\caption{Different speeds of motion}
	\label{fig:prim:speed}
	\end{subfigure}~
\caption{Primitives Layer evaluation. The magnitude parameter is discretized to two values: low (distance less than $0.5 \textrm{ft}$ and high for distance above $0.6 \textrm{ft}$ above the mobile device. We also discretize the speed to three values: high (edge duration $< 0.75$ sec), medium ($0.75$ sec $<$ edge duration between $<1.5$ sec), or low (edge duration $>1.5$ sec).}
\label{fig:primitives_layer}
\vspace{-0.2in}
\end{figure*}

In this section, we analyze the performance of \sys{} in two typical environments: (a) A typical apartment (Figure~\ref{fig:apartment_floorplan})
covering a 34$\times$39 $\textrm{ft}^2$ area and consisting of a living room connected to the dining area, three bedrooms, kitchen, and a bathroom. The walls have a thickness of $0.5 \textrm{ft}$  and the doors are made of wood and have a thickness of $0.16 \textrm{ft}$.
(b) The second floor of an engineering building at our campus (Figure~\ref{fig:vtmena_floorplan}) which contains 12 rooms connected by a large corridor and covers an area of 115$\times$49 $\textrm{ft}^2$.

In both environments, an HP EliteBook laptop is used as a receiver with a sampling rate of $50$ Hz controlled by the user using \sys{}. The apartment had two Cisco Linksys X2000 APs while the engineering building has three Netgear N300 APs installed in the experiment floor. We experimented with three different users producing more than 1000 primitive actions and hundreds of gestures including scenarios with seven interfering users in the same room, in addition to other people moving in the same floor during their daily life.

In addition to evaluating \sys{} at different locations uniformly distributed over the two environments, we extensively evaluate it in three different scenarios shown in Figure~\ref{fig:scenarios}: \textbf{1) No-wall/Line-of-sight}: The laptop and AP are in the same room. \textbf{2) Through-one-wall}: The laptop and AP are placed in adjacent rooms separated by one wall. \textbf{3) Through-two-walls}: The laptop and AP are placed in two different rooms separated by a corridor. Thus the signal penetrates two walls.

We first evaluate the accuracy of detecting primitives and gestures under these three scenarios with different realistic settings. Then we evaluate \sys{}'s performance at different locations uniformly distributed over both environments. 
 Unless otherwise specified, all results are averaged over both test environments.

\subsection{Primitives Layer Evaluation}

\subsubsection{Impact of Orientation}
We first evaluate edge detection performance at four orientations of the user relative to the device (East, West, North, and South) in the no-wall scenario (Figure~\ref{fig:scenarios}). The distance between the mobile device and the AP is set to $14 \textrm{ft}$. The results in Figure~\ref{fig:orientations} show that the overall primitive detection accuracy of \sys{} averaged over all different orientations is 90.5\%. The highest accuracy is achieved in the West orientation while the lowest is achieved in the East orientation. This is intuitive because in the East orientation, the human body blocks the line-of-sight between the device and the AP causing two negative effects: First, it produces noise in the received signal. Second, it reduces the received RSSI due to attenuation through the human body; stronger RSSI leads to better accuracy as we quantify later. We note that accuracy can be further enhanced by combining the RSSI from multiple APs as we show in the next section.

\subsubsection{Impact of Multiple APs}
Figure~\ref{fig:multiple_aps} shows the effect of using more than one AP on primitives detection accuracy averaged over different orientations. This experiment is performed in the engineering building due to the abundance of APs there. The overheard APs are over different floors in the building. A majority vote between all APs is used to select the correct primitive. The figure shows that increasing the number of APs to three increases the system accuracy to 96\%, and reaches 100\% when the device hears seven APs. Given the dense deployment of WiFi APs, this highlights that \sys{} can achieve robust performance in the many cases.

\subsubsection{Impact of Distance}
We investigate the relation between \sys{}'s accuracy and the coverage distance of an AP for the scenarios in Figure~\ref{fig:scenarios}. In this experiment, the distance between the device and AP varies and each primitive is repeated 25 times at each location for different orientations. Figure~\ref{fig:distance_gesture_detection} shows that the overall primitives detection accuracy decreases when increasing the distance between the device and AP. This is because stronger RSSIs lead to higher SNR and hence better accuracy. In other words, a weaker signal leads to lower changes in the signal strength in response to the hand movement, leading to less sensitivity. \sys{}, however, can still achieve more than $87\%$ accuracy for distances up to $26 \textrm{ft}$. This can be further enhanced by combining RSSIs from different APs as shown in the previous section.

\subsubsection{Individual primitives and attributes detection accuracy}

Finally, we evaluate the true positive (1- false negative) and false positive detection rate for the three primitives in addition to correctly identifying the primitive's  speed and distance parameters. The distance between the mobile device and the AP is set to $14 \textrm{ft}$.
The collected  dataset contains 350 samples of each primitive for a total of 1050 samples. Figure~\ref{fig:primitives_layer} shows the results. The figure shows that the different \sys{} processing modules lead to high detection accuracy with a high true positive rate and low false positive rate concurrently. In addition, \sys{} can detect the primitives' attributes with high accuracy, allowing it to be used in different applications. 

\begin{table}[!t]
    \centering
\scalebox{0.9}{
    \begin{tabular}{| l | p{0.3in} |p{0.3in} | p{0.3in} | p{0.3in} | p{0.3in} | p{0.3in} | }
    \hline
                  & Down-Up &Inf.& Up-Down &Up-Pause-Down& Down-Pause-Up&Un-known  \\ \hline \hline
       Down-Up    & \textbf{0.94}& 0   & 0   & 0   & 0   & 0.06\\ \hline
       Infinity   & 0.07   & \textbf{0.83}& 0.04   & 0   & 0.03   & 0.03\\ \hline
       Up-Down   & 0  & 0   & \textbf{1} & 0   & 0   & 0\\ \hline
       Up-Pause-Down & 0  & 0   & 0.05   & \textbf{0.9} & 0   & 0.05\\ \hline
       Down-Pause-Up & 0  & 0   & 0   & 0   & \textbf{0.96} &0.04\\ \hline
   \end{tabular}
}
    \caption{Confusion matrix for the different gesture families.}
    \label{tab:ges_fam}
\end{table}

\begin{figure}[!t]
\centering
	\begin{subfigure}[t]{0.23\textwidth}
		\centering
		\includegraphics[width=\textwidth]{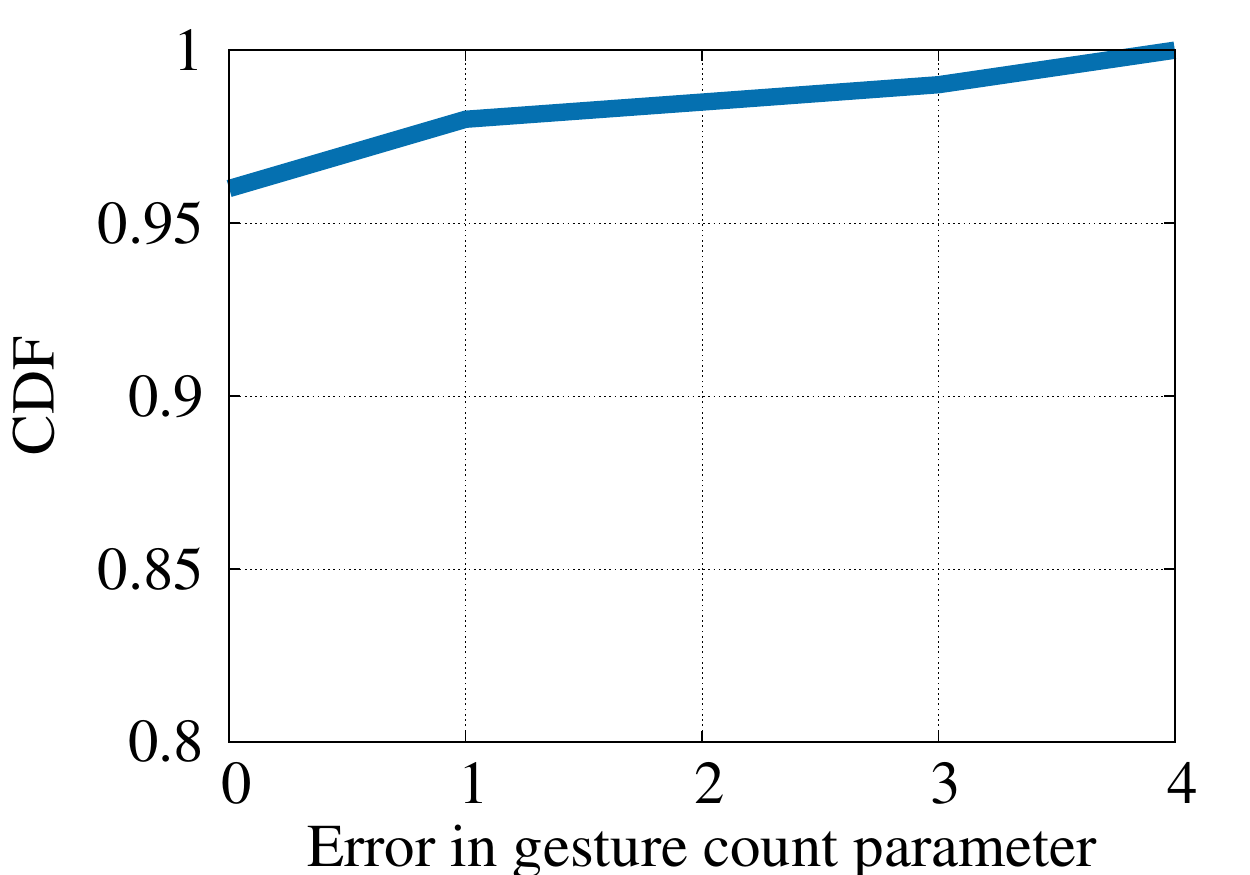}
		\caption{CDF of error in count parameter.}
		\label{fig:count}
	\end{subfigure}
	\begin{subfigure}[t]{0.23\textwidth}
		\centering
		\includegraphics[width=\textwidth]{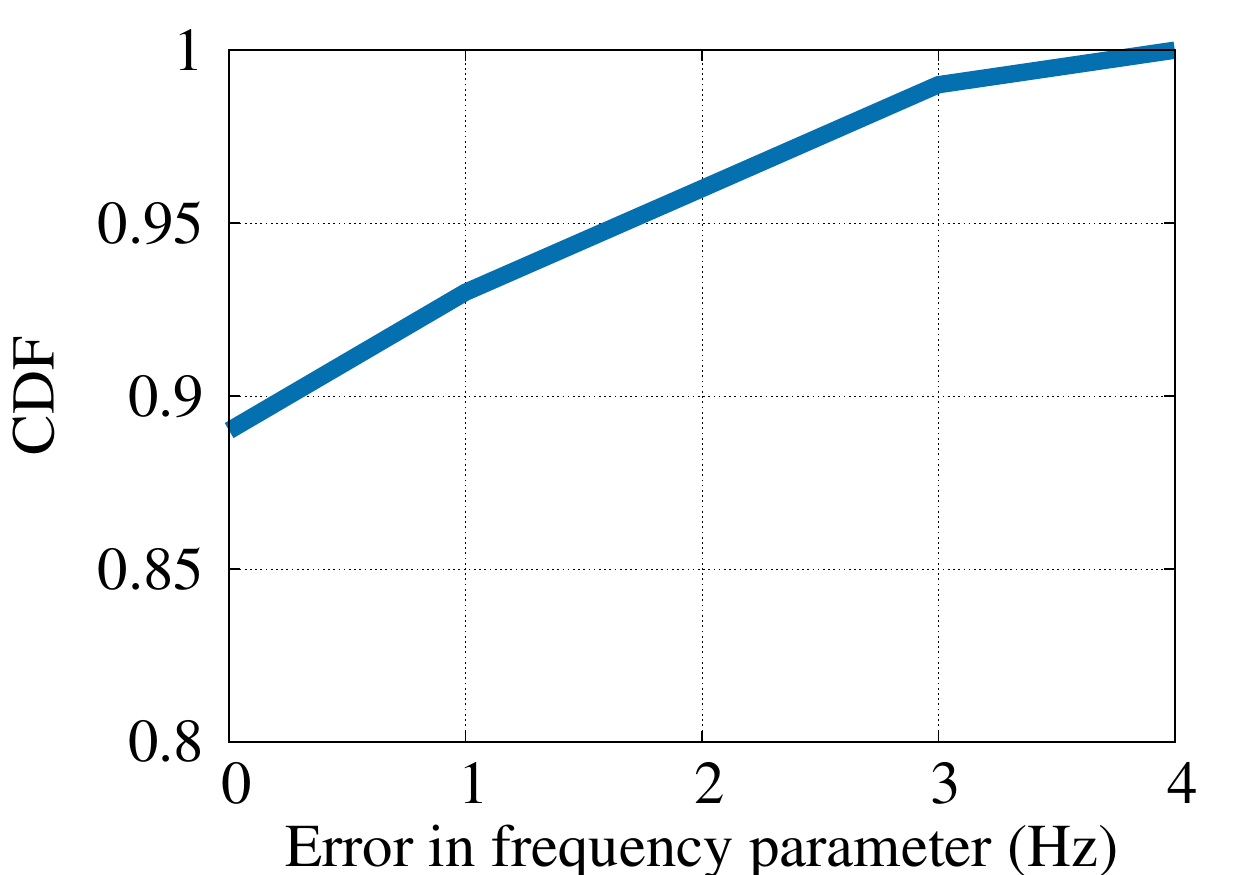}
		\caption{CDF of error in frequency parameter.}
		\label{fig:freq}
	\end{subfigure}
\caption{Gesture families parameters detection evaluation.}
\label{fig:count_freq}
\end{figure}

\subsection{Gesture Families Layer Evaluation}
We now evaluate the detection accuracy of the gesture families and their associated parameters (count and frequency). We consider the seven gesture families in Figure~\ref{fig:architecture:layers_overview}.

\subsubsection{Gesture detection}
A user performs a total of 192 gestures at a distance of $14 \textrm{ft}$ from the AP for the four different orientations relative to the mobile device using the AP the mobile device is associated with. The resulting confusion matrix is shown in Table~\ref{tab:ges_fam}. Overall, these results show that \sys{} can detect the gestures with a high accuracy of 92.6\%. This accuracy is higher than the primitives detection accuracy due to the processing we apply on gestures as described in Section~\ref{sec:pattern_enc}. This can be further increased by leveraging more APs.

\subsubsection{Gesture attributes detection accuracy}

Figure~\ref{fig:count_freq} shows the accuracy of detecting various gesture parameters. The figure shows that \sys{} can accurately detect the exact count 96\% of the time. This percentage increases to 98\% for a count error of one. Similarly, it can detect the repetition frequency of a gesture within one second 93\% of the time.

\subsection{Whole-home Gesture Recognition Case Study}
We evaluate the system in our apartment environment using a multi-media player application. Both APs installed in the apartment are used, and since there is no majority vote here, the AP with the strongest signal is used. We evaluate the multi-media player action detection performance at eight locations uniformly distributed over the apartment and covering different scenarios (line-of-sight, through-one-wall, and through-two-walls). At each location, the user performs the seven actions shown in Figure~\ref{fig:architecture:layers_overview}. Each action is performed 20 times at each location for a total of 1120 actions. Table~\ref{tab:app_confusion} shows the confusion matrix for the seven gestures across all locations. The matrix shows that the overall classification accuracy over all application actions is 96\%. This accuracy shows \sys{}'s ability to use wireless signals to extract rich gestures.

\subsection{Gesture Recognition with Interference}
We now evaluate \sys{} in the presence interference due to noise: in the absence of gestures, and as a result of the presence of other humans.

\begin{table}[!t]
\centering
\includegraphics[width=3in]{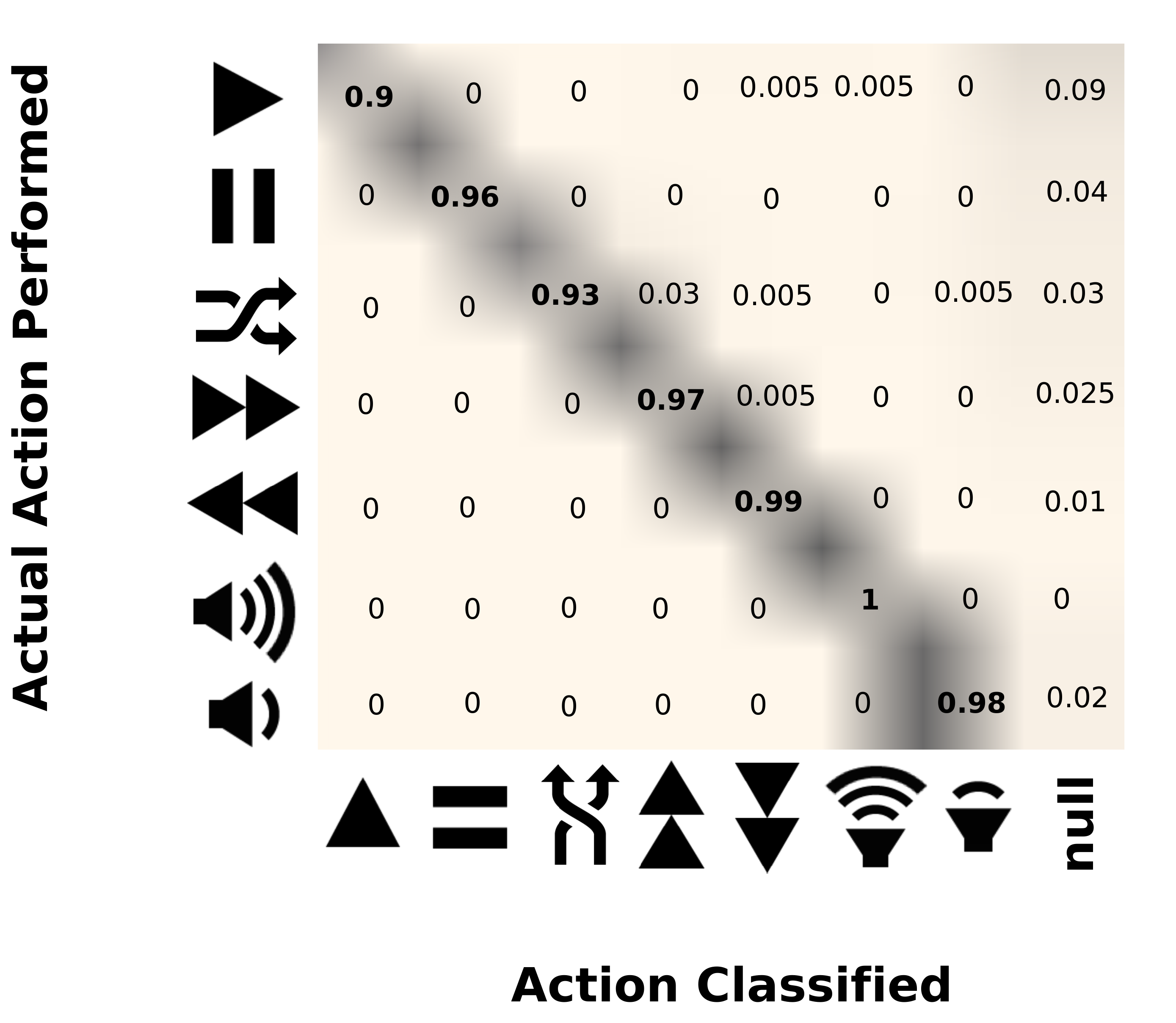}
\caption{Confusion matrix for the media player application actions.}
\label{tab:app_confusion}
\vspace{-0.2in}
\end{table}

\subsubsection{False detection rate in the absence of actual gestures}
As discussed before, \sys{} uses a unique preamble with a repetitive pattern to compensate for the environment noise. We perform an experiment in the lab in our engineering building (Figure~\ref{fig:testbeds}) to detect the false detection rate in the absence of actual gestures. For this, the laptop is left running from 9:00am to 5:00pm. The room contains six students performing their daily routine as well as a dynamic number of visitors during the day. The maximum number of interfering humans during the experiment is seven.

\begin{figure}[!t]
\centering
\includegraphics[width=3in]{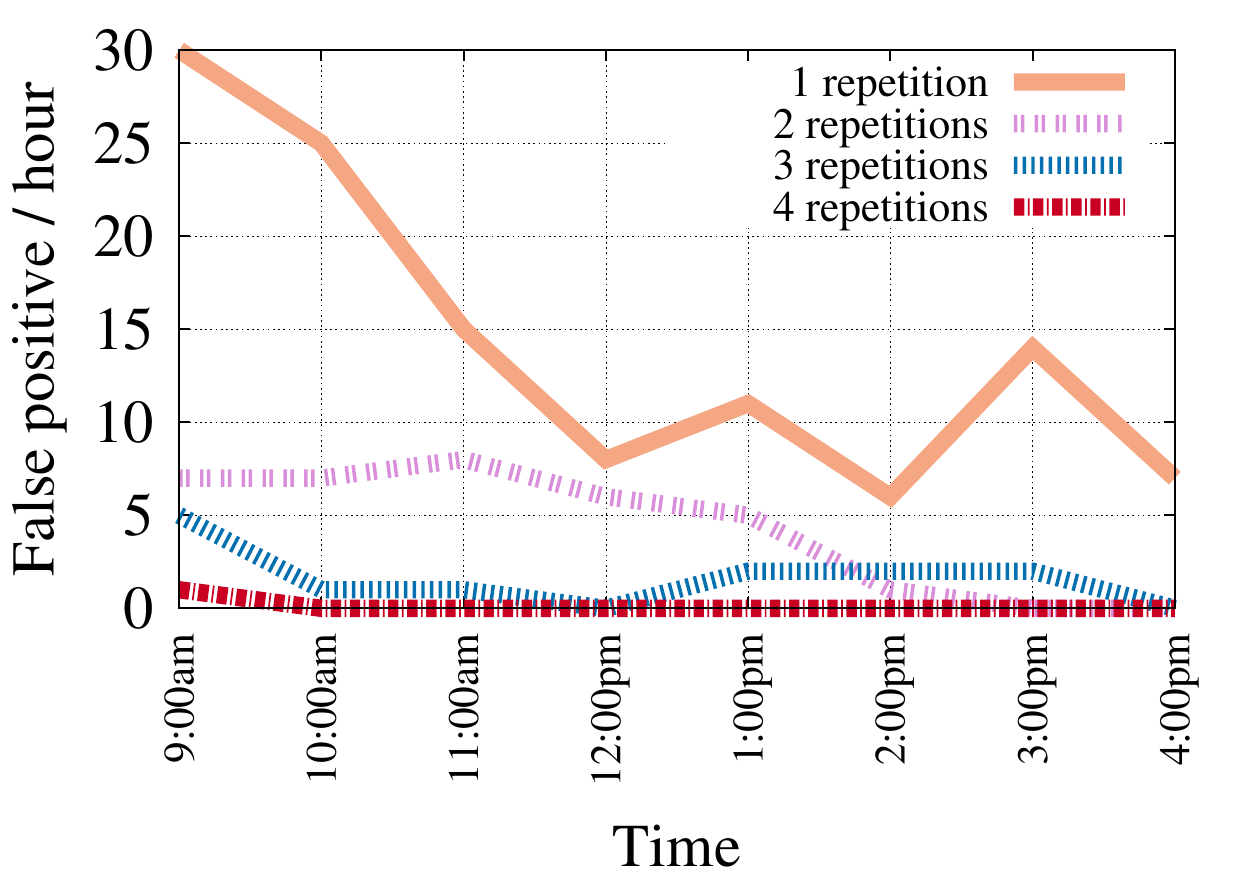}
\caption{False detection rate in the absence of actual gestures.}
\label{fig:false_preamble}
\vspace{-0.2in}
\end{figure}

Figure~\ref{fig:false_preamble} shows the false detection rate, averaged per hour. We observe that as the repetition within the preamble increases, accuracy increases. A preamble length of two (used throughout the paper) is enough to obtain an average of 3.7 false preamble detections per hour. This can be further reduced to about zero false detections per hour by increasing the preamble length to four.

\subsubsection{Gestures detection accuracy in the presence of other humans}
In this experiment, the distance between the AP and the receiver is 23 $\textrm{ft}$. Our experiment has up to four interfering users at random locations in the same room performing random actions. Figure~\ref{fig:humans} shows the overall gesture detection accuracy as we increase the number of interfering humans. The figure shows that, even with one AP, the accuracy is as high as 89\% in the presence of four interfering users. The reason for this accuracy is due to the closer distance between the \sys{} actual user and device compared to the interferers as we quantify in the next section. We note, however, that increasing the number of interfering users within a small area, e.g. a conference scenario, may affect the system accuracy. This needs to be further investigated. 

\subsubsection{Distance from the interfering human}
\begin{figure}[!t]
\centering
	\begin{subfigure}[b]{0.23\textwidth}
		\centering
		\includegraphics[width=\textwidth]{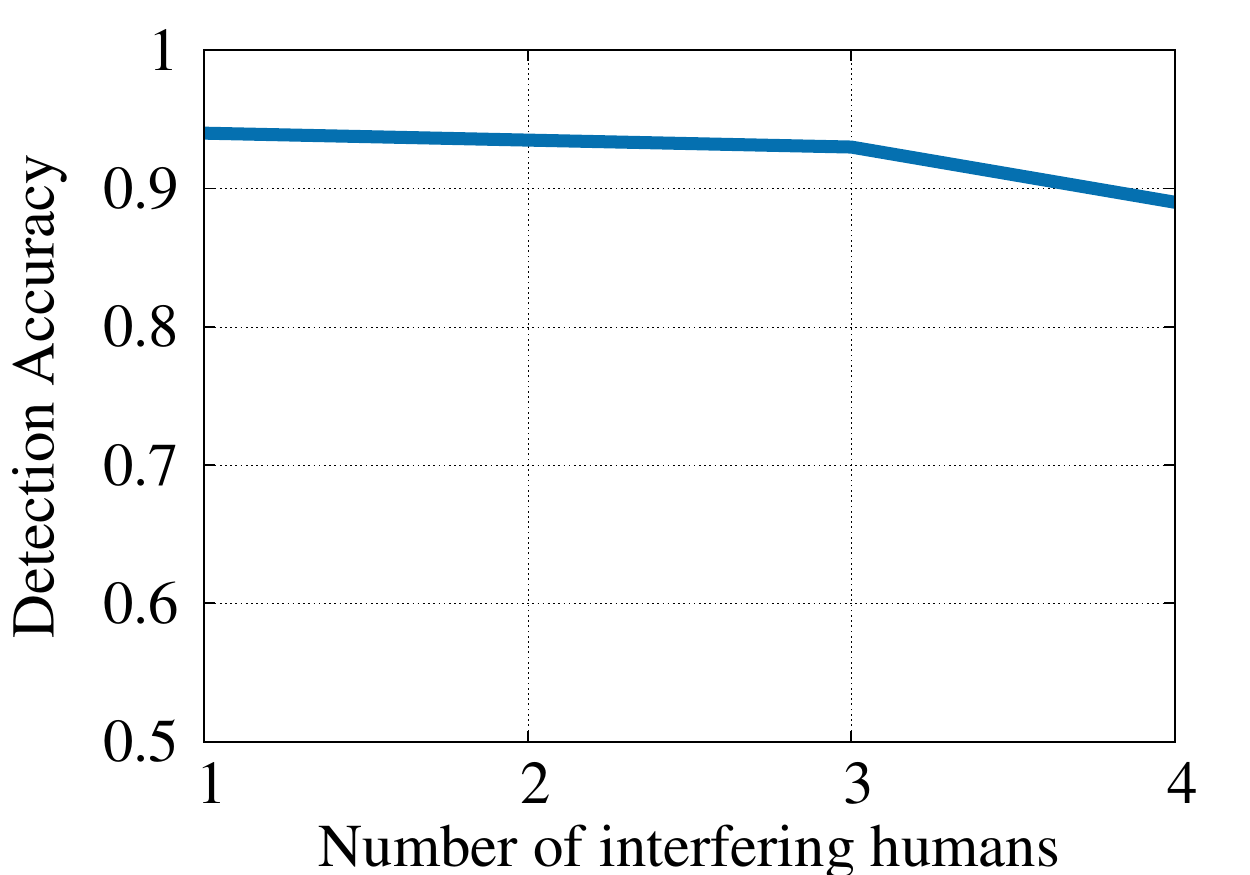}
		\caption{Different number of interfering humans.}
		\label{fig:humans}
	\end{subfigure}~
	\begin{subfigure}[b]{0.23\textwidth}
		\centering
		\includegraphics[width=\textwidth]{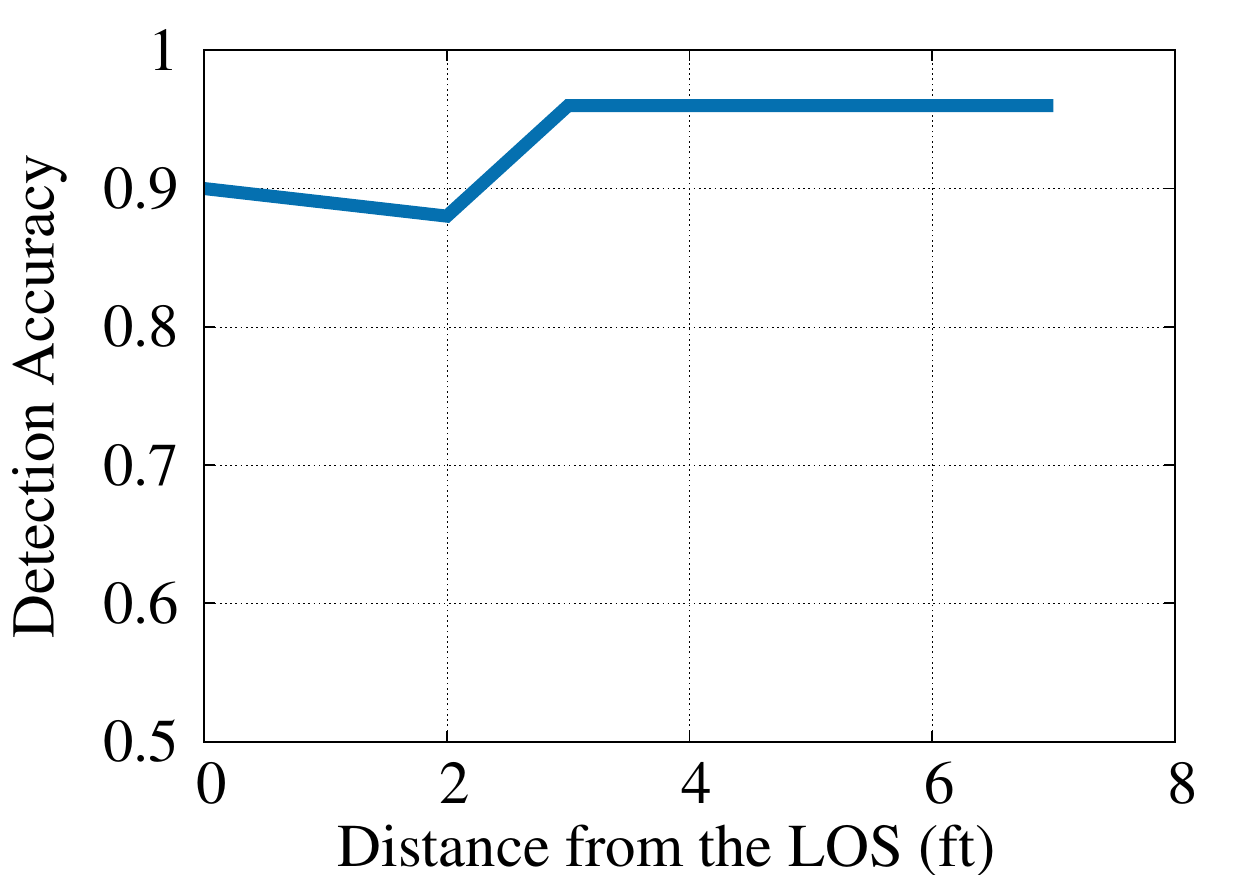}
		\caption{One human with different distances from the LOS.}
		\label{fig:stress}
	\end{subfigure}
\caption{Effect of interfering humans on \sys{} accuracy.}
\vspace{-0.2in}
\label{fig:interf_humans}
\end{figure}

Since \sys{} depends on the changes in RSSI, humans affecting the LOS between the transmitter and receiver may severely affect the detection accuracy. To evaluate \sys{}'s gesture classification
accuracy, we perform an experiment where the distance between the AP and the device is set to $14 \textrm{ft}$ and an interfering user moves in different positions along the line perpendicular to the line joining the transmitter and receiver. Figure~\ref{fig:stress} reflects the results, showing that as long as the interfering user is more than $4 \textrm{ft}$ away, she has no effect on accuracy. However, a close-by user by less than $3 \textrm{ft}$ reduces the system accuracy. This, however, is mitigated by the diversity of APs as well as the unique preamble used by \sys{}.

\section{Conclusion}
\label{conclusion}
We presented \sys{}, a robust gesture recognition system that uses WiFi RSSI's to detect human hand motions around a user device. \sys{} does not require any modification to the available wireless equipment or any extra sensors, and does not require any training prior to deployment. We addressed system challenges including signal denoising, gesture primitives extraction, reducing the false positive rate of gesture detection, and adapting to changing signal polarity caused by environmental interference. \sys{}'s energy-efficiency stems from using a preamble that is easy to detect based on a simple thresholding approach as well as using the wavelet transform that works in linear time in its various processing modules.

Extensive evaluation of \sys{} in two environments shows that it can detect basic primitives with an accuracy of 87.5\% using a single AP for distances up to 26 \textit{ft} including through-the-wall non-line-of-sight scenarios. This accuracy increases to 96\% using three overheard APs, which is the typical case for many urban WiFi scenarios. In addition, \sys{} can achieve a classification accuracy of 96\% for the application actions. These results are robust to the presence of other interfering humans, highlighting \sys{}'s ability for enabling future ubiquitous hands-free gesture-based interaction with mobile devices.

Currently, we are extending the system in multiple directions including leveraging detailed channel state information (CSI) from the physical layer to further enhance accuracy and generate finer grained gesture families, leveraging other ubiquitous wireless technologies, such as cellular and bluetooth, among others.

\bibliographystyle{IEEEtran}

\begin{thebibliography}{10}
\providecommand{\url}[1]{#1}
\csname url@samestyle\endcsname
\providecommand{\newblock}{\relax}
\providecommand{\bibinfo}[2]{#2}
\providecommand{\BIBentrySTDinterwordspacing}{\spaceskip=0pt\relax}
\providecommand{\BIBentryALTinterwordstretchfactor}{4}
\providecommand{\BIBentryALTinterwordspacing}{\spaceskip=\fontdimen2\font plus
\BIBentryALTinterwordstretchfactor\fontdimen3\font minus
  \fontdimen4\font\relax}
\providecommand{\BIBforeignlanguage}[2]{{%
\expandafter\ifx\csname l@#1\endcsname\relax
\typeout{** WARNING: IEEEtran.bst: No hyphenation pattern has been}%
\typeout{** loaded for the language `#1'. Using the pattern for}%
\typeout{** the default language instead.}%
\else
\language=\csname l@#1\endcsname
\fi
#2}}
\providecommand{\BIBdecl}{\relax}
\BIBdecl

\bibitem{gupta2012soundwave}
S.~Gupta, D.~Morris, S.~Patel, and D.~Tan, ``Soundwave: using the doppler
  effect to sense gestures,'' in \emph{Proceedings of the 2012 ACM annual
  conference on Human Factors in Computing Systems}.\hskip 1em plus 0.5em minus
  0.4em\relax ACM, 2012, pp. 1911--1914.

\bibitem{QosmioG55}
R.~Block, ``Toshiba {Q}osmio {G}55 features {S}purs{E}ngine, visual gesture
  controls,''
  {http://www.engadget.com/2008/06/14/toshiba-qosmio-g55-features-spursengine-visual-gesture-controls/},
  last accessed March 1st, 2014.

\bibitem{RakuNaviGPS}
A.~Santos, ``Pioneer's latest {R}aku {N}avi {GPS} units take commands from hand
  gestures,''
  {http://www.engadget.com/2012/10/07/pioneer-raku-navi-gps-hand-gesture-controlled/},
  last accessed March 1st, 2014.

\bibitem{shotton2013real}
J.~Shotton, T.~Sharp, A.~Kipman, A.~Fitzgibbon, M.~Finocchio, A.~Blake,
  M.~Cook, and R.~Moore, ``Real-time human pose recognition in parts from
  single depth images,'' \emph{Communications of the ACM}, vol.~56, no.~1, pp.
  116--124, 2013.

\bibitem{cohn2012humantenna}
G.~Cohn, D.~Morris, S.~Patel, and D.~Tan, ``Humantenna: using the body as an
  antenna for real-time whole-body interaction,'' in \emph{Proceedings of the
  2012 ACM annual conference on Human Factors in Computing Systems}.\hskip 1em
  plus 0.5em minus 0.4em\relax ACM, 2012, pp. 1901--1910.

\bibitem{harrison2010skinput}
C.~Harrison, D.~Tan, and D.~Morris, ``Skinput: appropriating the body as an
  input surface,'' in \emph{Proceedings of the SIGCHI Conference on Human
  Factors in Computing Systems}.\hskip 1em plus 0.5em minus 0.4em\relax ACM,
  2010, pp. 453--462.

\bibitem{kim2012digits}
D.~Kim, O.~Hilliges, S.~Izadi, A.~D. Butler, J.~Chen, I.~Oikonomidis, and
  P.~Olivier, ``Digits: freehand 3d interactions anywhere using a wrist-worn
  gloveless sensor,'' in \emph{Proceedings of the 25th annual ACM symposium on
  User interface software and technology}.\hskip 1em plus 0.5em minus
  0.4em\relax ACM, 2012, pp. 167--176.

\bibitem{scholz2011challenges}
M.~Scholz, S.~Sigg, H.~R. Schmidtke, and M.~Beigl, ``Challenges for device-free
  radio-based activity recognition,'' in \emph{Workshop on Context Systems,
  Design, Evaluation and Optimisation}, 2011.

\bibitem{adib2013see}
F.~Adib and D.~Katabi, ``See through walls with wifi!'' in \emph{Proceedings of
  the ACM SIGCOMM}.\hskip 1em plus 0.5em minus 0.4em\relax ACM, 2013, pp.
  75--86.

\bibitem{pu2013whole}
Q.~Pu, S.~Gupta, S.~Gollakota, and S.~Patel, ``Whole-home gesture recognition
  using wireless signals,'' in \emph{Proceedings of the 19th annual
  international conference on Mobile computing \& networking}.\hskip 1em plus
  0.5em minus 0.4em\relax ACM, 2013, pp. 27--38.

\bibitem{kosba2012rasid}
A.~E. Kosba, A.~Saeed, and M.~Youssef, ``Rasid: A robust {WLAN} device-free
  passive motion detection system,'' in \emph{Pervasive Computing and
  Communications (PerCom), 2012 IEEE International Conference on}.\hskip 1em
  plus 0.5em minus 0.4em\relax IEEE, 2012, pp. 180--189.

\bibitem{youssef2007challenges}
M.~Youssef, M.~Mah, and A.~Agrawala, ``Challenges: device-free passive
  localization for wireless environments,'' in \emph{Proceedings of the 13th
  annual ACM international conference on Mobile computing and
  networking}.\hskip 1em plus 0.5em minus 0.4em\relax ACM, 2007, pp. 222--229.

\bibitem{saeed2014ichnaea}
A.~Saeed, A.~E. Kosba, and M.~Youssef, ``Ichnaea: A low-overhead robust {WLAN}
  device-free passive localization system,'' \emph{IEEE Journal of Selected
  Topics in Signal Processing}, vol.~8, no.~1, pp. 5--15, 2014.

\bibitem{abdel2013monophy}
H.~Abdel-Nasser, R.~Samir, I.~Sabek, and M.~Youssef, ``{MonoPHY}:
  Mono-stream-based device-free {WLAN} localization via physical layer
  information,'' in \emph{Wireless Communications and Networking Conference
  (WCNC), 2013 IEEE}.\hskip 1em plus 0.5em minus 0.4em\relax IEEE, 2013, pp.
  4546--4551.

\bibitem{abdelnasser2015wigest}
H.~Abdelnasser, K.~Harras, and M.~Youssef, ``{WiGest Demo}:
   A Ubiquitous WiFi-based Gesture Recognition System'' in
   \emph{INFOCOM demo 2015}.\relax IEEE, 2015.

\bibitem{Kafrawy:PIMRC11}
K.~El-Kafrawy, M.~Youssef, and A.~El-Keyi, ``Impact of the human motion on the
  variance of the received signal strength of wireless links,'' in
  \emph{PIMRC}, 2011, pp. 1208--1212.

\bibitem{aly2013new}
H.~Aly and M.~Youssef, ``New insights into wifi-based device-free
  localization,'' in \emph{Proceedings of the 2013 ACM conference on Pervasive
  and ubiquitous computing adjunct publication}.\hskip 1em plus 0.5em minus
  0.4em\relax ACM, 2013, pp. 541--548.

\bibitem{seifeldin2011kalman}
M.~A. Seifeldin, A.~F. El-keyi, and M.~A. Youssef, ``Kalman filter-based
  tracking of a device-free passive entity in wireless environments,'' in
  \emph{ACM international workshop on Wireless network testbeds, experimental
  evaluation and characterization}, 2011.

\bibitem{seifeldin2010deterministic}
M.~Seifeldin and M.~Youssef, ``A deterministic large-scale device-free passive
  localization system for wireless environments,'' in \emph{Proceedings of the
  3rd International Conference on PErvasive Technologies Related to Assistive
  Environments}.\hskip 1em plus 0.5em minus 0.4em\relax ACM, 2010, p.~51.

\bibitem{eleryan2011aroma}
A.~Eleryan, M.~Elsabagh, and M.~Youssef, ``Synthetic generation of radio maps
  for device-free passive localization,'' in \emph{IEEE Global
  Telecommunications Conference (GLOBECOM 2011)}.

\bibitem{kosba2012robust}
A.~E. Kosba, A.~Saeed, and M.~Youssef, ``Robust {WLAN} device-free passive
  motion detection,'' in \emph{IEEE Wireless Communications and Networking
  Conference (WCNC 2012)}, pp. 3284--3289.

\bibitem{sabek2012multi}
I.~Sabek and M.~Youssef, ``Multi-entity device-free {WLAN} localization,'' in
  \emph{IEEE GLOBECOM 2012}.

\bibitem{smartdevices}
M.~Moussa and M.~Youssef, ``Smart devices for smart environments: Device-free
  passive detection in real environments,'' in \emph{IEEE Percom Workshops
  2009.}, 2009.

\bibitem{xiao2013pilot}
J.~Xiao, K.~Wu, Y.~Yi, L.~Wang, and L.~M. Ni, ``Pilot: Passive device-free
  indoor localization using channel state information,'' in \emph{Distributed
  Computing Systems (ICDCS), 2013 IEEE 33rd International Conference on}.\hskip
  1em plus 0.5em minus 0.4em\relax IEEE, 2013, pp. 236--245.

\bibitem{kosba2009analysis}
A.~E. Kosba, A.~Abdelkader, and M.~Youssef, ``Analysis of a device-free passive
  tracking system in typical wireless environments,'' in \emph{New
  Technologies, Mobility and Security (NTMS), 2009 3rd International Conference
  on}.\hskip 1em plus 0.5em minus 0.4em\relax IEEE, 2009, pp. 1--5.

\bibitem{el2010propagation}
K.~El-Kafrawy, M.~Youssef, A.~El-Keyi, and A.~Naguib, ``Propagation modeling
  for accurate indoor {WLAN} {RSS}-based localization,'' in \emph{IEEE
  Vehicular Technology Conference Fall (VTC 2010-Fall)}.

\bibitem{seifeldin2013nuzzer}
M.~Seifeldin, A.~Saeed, A.~E. Kosba, A.~El-Keyi, and M.~Youssef, ``Nuzzer: A
  large-scale device-free passive localization system for wireless
  environments,'' \emph{Mobile Computing, IEEE Transactions on}, vol.~12,
  no.~7, pp. 1321--1334, 2013.

\bibitem{Sabek:TRSPOT12}
I.~Sabek and M.~Youssef, ``Spot: An accurate and efficient multi-entity
  device-free {WLAN} localization system,'' \emph{CoRR}, vol. abs/1207.4265,
  2012.

\bibitem{sabek2014ace}
I.~Sabek, M.~Youssef, and A.~Vasilakos, ``{ACE}: An accurate and efficient
  multi-entity device-free {WLAN} localization system,'' \emph{IEEE
  Transactions on Mobile Computing}, 2014.

\bibitem{ding2011rftraffic}
Y.~Ding, B.~Banitalebi, T.~Miyaki, and M.~Beigl, ``{RFTraffic}: passive traffic
  awareness based on emitted {RF} noise from the vehicles,'' in \emph{IEEE 2011
  ITS Telecommunications Conference}.

\bibitem{al2012rf}
A.~Al-Husseiny and M.~Youssef, ``{RF}-based traffic detection and
  identification,'' in \emph{IEEE Vehicular Technology Conference (VTC Fall),
  2012}, 2012.

\bibitem{sigg2013rfa}
S.~Sigg, S.~Shi, and Y.~Ji, ``{RF}-based device-free recognition of
  simultaneously conducted activities,'' in \emph{Proceedings of the 2013 ACM
  conference on Pervasive and ubiquitous computing adjunct publication}.\hskip
  1em plus 0.5em minus 0.4em\relax ACM, 2013, pp. 531--540.

\bibitem{Kassem:VTC12}
N.~Kassem, A.~E. Kosba, and M.~Youssef, ``{RF}-based vehicle detection and
  speed estimation,'' in \emph{IEEE VTC Spring}, 2012.

\bibitem{shi2013joint}
S.~Shi, S.~Sigg, and Y.~Ji, ``Joint localization and activity recognition from
  ambient fm broadcast signals,'' in \emph{Proceedings of the 2013 ACM
  conference on Pervasive and ubiquitous computing adjunct publication}.\hskip
  1em plus 0.5em minus 0.4em\relax ACM, 2013, pp. 521--530.

\bibitem{alzantot2012crowdinside}
M.~Alzantot and M.~Youssef, ``{CrowdInside}: automatic construction of indoor
  floorplans,'' in \emph{ACM 2012 SIGSpatial GIS Conference}.

\bibitem{checkinside}
M.~ELhamshary and M.~Youssef, ``{CheckInside}: An indoor location-based social
  network.'' in \emph{ACM Ubicomp 2014}.

\bibitem{wang2012no}
H.~Wang, S.~Sen, A.~Elgohary, M.~Farid, M.~Youssef, and R.~R. Choudhury, ``No
  need to war-drive: unsupervised indoor localization,'' in \emph{ACM MobiSys
  2012}.

\bibitem{aumi2013doplink}
M.~T.~I. Aumi, S.~Gupta, M.~Goel, E.~Larson, and S.~Patel, ``Doplink: using the
  doppler effect for multi-device interaction,'' in \emph{Proceedings of the
  2013 ACM international joint conference on Pervasive and ubiquitous
  computing}.\hskip 1em plus 0.5em minus 0.4em\relax ACM, 2013, pp. 583--586.

\bibitem{kim2009human}
Y.~Kim and H.~Ling, ``Human activity classification based on micro-doppler
  signatures using a support vector machine,'' \emph{Geoscience and Remote
  Sensing, IEEE Transactions on}, vol.~47, no.~5, pp. 1328--1337, 2009.

\bibitem{Wavlet_book}
S.~Mallat, \emph{A Wavelet Tour of Signal Processing, Third Edition: The Sparse
  Way}, 3rd~ed.\hskip 1em plus 0.5em minus 0.4em\relax Academic Press, 2008.

\bibitem{sardy2001robust}
S.~Sardy, P.~Tseng, and A.~Bruce, ``Robust wavelet denoising,'' \emph{Signal
  Processing, IEEE Transactions on}, vol.~49, no.~6, pp. 1146--1152, 2001.

\bibitem{wavlet_adv}
S.~Nibhanupudi, ``Signal denoising using wavelets,'' Master's thesis,
  University of Cincinnati, 2003.

\end{thebibliography}

\end{document}